\documentclass[twocolumn,showpacs,preprintnumbers,amsmath,amssymb,pra,floatfix]{revtex4}
\usepackage{graphicx}% Include figure files
\usepackage{dcolumn}% Align table columns on decimal point
\usepackage{bm}% bold math 

\begin{document}
\preprint{  }
%
%
%========= TITLE =====================================================

\title{Resonant effects on the two-photon emission from hydrogenic ions}

%\titlerunning{Resonant effects on the two-photon emission from
%hydrogenic ions} 

%
%========= AUTHORS ===================================================
%
%
\author{P. Amaro}
\email{p.d.g.amaro@gmail.com}
\author{J.\ P.\ Santos}
\email{jps@fct.unl.pt}
\author{ F.\ Parente}
\email{facp@fct.unl.pt}
\affiliation{CFA, Departamento de F{\'\i}sica, Faculdade de Ci{\^e}ncias e
  Tecnologia, \\
  FCT, Universidade Nova de Lisboa, 2829-516 Caparica, Portugal
}
\author{A.\ Surzhykov} \email{surz@physi.uni-heidelberg.de}
\affiliation{Physikalisches Institut, Universit\"at Heidelberg,
  Philosophenweg 12, D-69120 Heidelberg, Germany \\ and GSI
  Helmholtzzentrum f\"ur Schwerionenforschung GmbH, Planckstrasse 1,
  D-64291 Darmstadt, Germany}
\author{P.\ Indelicato}
\email{paul.indelicato@spectro.jussieu.fr}
\affiliation{Laboratoire Kastler Brossel, \'Ecole Normale Sup\' erieure; 
  CNRS; Universit\' e P. et M. Curie - Paris 6, Case 74; 4, place
  Jussieu, 75252 Paris CEDEX 05, France
}
  
\date{\today}% It is always \today, today,
%========= ABSTRACT ==================================================

\begin{abstract}
  A theoretical study the all two-photon transitions from initial
  bound states with $n_i=2,3$ in hydrogenic ions is presented.
  High-precision values of relativistic decay rates for ions with
  nuclear charge in the range $1\le Z \le 92$ are obtained through the
  use of finite basis sets for the Dirac equation constructed from
  B-splines.  We also report the spectral (energy) distributions of
  several resonant transitions, which exhibit interesting structures,
  such as zeroes in the emission spectrum, indicating that two-photon
  emission is strongly suppressed at certain frequencies. We compare
  two different approaches (the Line Profile Approach (LPA) and the
  QED approach based on the analysis of the relativistic two-loop self
  energy (TLA)) to regularize the resonant contribution to the decay
  rate. Predictions for the pure two-photon contributions obtained in
  these approaches are found to be in a good numerical agreement.
\end{abstract}

%
%end of abstract
%
%========= CLASSIFICATION ============================================
%
%\pacs{
%      {31.30.Jv}{Relativistic and quantum electrodynamics effects
%      in atoms and molecules}   \and
%      {32.70.Fw}{Absolute and relative intensities}   \and
%      {32.80.Wr}{Other multiphoton processes}
%     } % end of PACS codes
%
\maketitle
%
%
%
%=====================================================================
%
\section{\label{intro}Introduction}
%\label{intro}

Two-photon transition in hydrogen and hydrogenlike ions are under
investigation since G\"oeppert-Mayer presented her theoretical
formalism in 1931 \cite{162}. The early interest on these transitions
from metastable states of hydrogen came mainly from astrophysics
\cite{159,2276}, which was recently revived by Chluba and Sunyaev
\cite{1417}.
Among many applications of recent two-photon studies,
one can cite the determination of the Rydberg constant
\cite{2277,2278,2896}, measurement of the Lamb-shift \cite{2278,142},
testing Bell's inequality \cite{2279}, as well as various applications
in molecular spectroscopy \cite{2826}, tissue imaging \cite{2857} and
protein structure analysis \cite{2827}.  Another interest in
two-photon transitions is connected to the study of parity-violation
effects in H-like and He-like ions \cite{1913,65}.  
%Recently, Chluba
%and Sunyaev \cite{1417} revived the astrophysics interest in
%two-photon transitions.
%
The two-photon spectral distribution has recently been used for
precise efficiency calibration of solid-state X-rays detector as it
has a known shape for a large distribution of energies \cite{2897}.

Similar to single-photon processes, two-photon emission can be
spontaneous or stimulated, whereas two-photon absorption is only
stimulated. However, since each photon carries one unit of angular
momentum in the dipole approximation, certain transitions between
atomic energy levels, forbidden as single-photon processes, are
allowed as two-photon processes.  Another important distinction lies
in the fact that the emission spectrum of spontaneous two-photon
transitions is continuous unlike the spectrum in a single-photon
process.  A continuous spectrum is possible because energy
conservation requires only that the sum of both photon energies equals
the energy of the transition. For the transition
\begin{equation}
\left( n_{i},j_{i} \right) \rightarrow \left(n_{f},j_{f} \right)+\hbar
\omega _{1}+\hbar \omega _{2}, 
\label{n_transi}
\end{equation}
where $\left( n_{i},j_{i} \right)$ and $\left(n_{f},j_{f} \right)$
denote the principal quantum numbers and total angular momenta of the
initial and final hydrogenic states, respectively, and $\hbar \omega
_{1}$ and $\hbar \omega _{2}$ are the energies of each photon, the
conservation of the energy leads to the condition
\begin{equation}
E_{f}-E_{i}= \hbar \omega _{1}+ \hbar \omega _{2},  
\label{con_ener}
\end{equation}
where $E_{i}$ and $E_{f}$ are the energies of the initial and final
ionic states, respectively.

Because of its importance, the $2s_{1/2}\rightarrow 1s_{1/2}$
two-photon transition rate in hydrogen has been calculated and
discussed many times using different approaches. An historical
overview from both theoretical and experimental point of view can be
found in the 1998 article by Santos {\it et al} \cite{388}.

Recently, Surzhykov {\it et al} \cite{1353} performed a relativistic
calculation to study the angular correlations in the two-photon decay
of hydrogenlike ions and Labzowsky $et$ $al$ \cite{1418} evaluated the
$2E1$ contribution for the $2s_{1/2}\rightarrow 1s_{1/2}$ transition
and the $E1M1$ and $E1E2$ contributions for the $2p_{1/2}\rightarrow
1s_{1/2}$ transition using an expression similar to the one obtained
by Goldman and Drake \cite{3} in the Quantum Electrodynamics (QED)
framework.  Also in this framework, Nganso {\it et al} \cite{1919}
carried out the treatment of the $S$ matrix for bound-bound
transitions.

In this work, which uses techniques of a previous one \cite{388}, we
study the two--photon decay of several excited states using two
approaches to deal with resonances; the Line Profile Approach (LPA)
\cite{1434} and the QED approach based on the analysis of the
relativistic two-loop self energy (TLA) to regularize the resonant
contribution to the decay rate~\cite{2285,2491}. We present calculated
values for two--photon decay rates obtained with both approaches for
one--electron ions with a nuclear charge up to 92.  This article is
organized as follows: in Sec.  \ref{theory} we give a brief review of
the background theory involved in two--photon emission, in Sec.
\ref{Numer_result} we present the results obtained in this work and
the conclusions are presented in Sec. \ref{Con}.
%
%
%=====================================================================
%
\section{\label{theory}Theory of relativistic radiative transitions}
%\label{theory}

%
%---------------------------------------------------------------------
%
\subsection{Two-photon spontaneous emission}

%
%.....................................................................
%
\subsubsection{\label{formalism}General formalism}
%\label{theory_nonresonant}

The relativistic theory of two-photon transitions is given in detail
in references \cite{3,12,388}. 

In the present work, therefore, we report only the most important
equations and notations used.  

The basic expression for the differential (in energy of one of the
photons) rate is, in atomic units,
\begin{eqnarray}
\frac{dw}{d\omega _{1}} &=&\frac{\omega _{1}\omega _{2}}{(2\pi )^{3}c^{2}}%
\left\vert \sum_{\nu} \left( \frac{\left\langle f\left\vert A_{2}^{\ast }\right\vert
\nu \right\rangle \left\langle \nu \left\vert A_{1}^{\ast }\right\vert
i\right\rangle }{E_{\nu}-E_{i}+\omega _{1}}\right.\right.  \nonumber \\
&&\left. \left.+\frac{\left\langle f\left\vert A_{1}^{\ast }\right\vert
\nu \right\rangle \left\langle \nu \left\vert A_{2}^{\ast }\right\vert
i\right\rangle }{E_{\nu}-E_{i}+\omega _{2}}\right) \right\vert ^{2}d\Omega _{1}d\Omega
_{2},  
\label{eq_gera}
\end{eqnarray}%
where $\omega _{j}$ is the frequency and $d\Omega _{j}$ is the element
of solid angle of the $j^{th}$ photon, and $c$ is the speed of light.
The frequencies of the photons are constrained by the energy
conservation Eq. (\ref{con_ener}).

For photon plane-wave with propagation vector $\mbox{\boldmath$k$}_j$
and polarization vector $\hat{\mbox{\boldmath$e$}}_j$ ($\hat{\mbox{\boldmath$e$}}_j
\cdot \mbox{\boldmath$k$}_j$=0), the operators $A_{j}^{*}$
in (\ref{eq_gera}) are given by 
\begin{equation}
       A_{j}^{*}= \mbox{\boldmath$\alpha$} \cdot
       \left( \hat{\mbox{\boldmath$e$}}_j +G \hat{\mbox{\boldmath$k$}}_j \right) 
       e^{-i \mbox{\boldmath$k$}_j \cdot \mbox{\boldmath$r$}}
       - G e^{-i \mbox{\boldmath$k$}_j \cdot \mbox{\boldmath$r$}}
\label{2.3}
\end{equation}
where $\mbox{\boldmath$\alpha$}$ are Dirac matrices and $G$ is an
arbitrary gauge parameter.  Among the large variety of possible
gauges, Grant \cite{6} showed that there are two values of $G$ which
are of particular utility because they lead to well--known
nonrelativistic operators. If $G=0$, one has the so called Coulomb
gauge, or velocity gauge, which leads to the dipole velocity form in
the nonrelativistic limit.  If $G=\left[ \left( L+1\right) /L\right]
^{1/2}$, for example, $G=\sqrt{2}$ for $E1$ transitions ($L=1$), one
obtains a nonrelativistic expression which reduces to the dipole
length form of the transition operator.  The two-photon transitions
gauge invariance was studied by Goldman and Drake \cite{3}. From the
general requirement of gauge invariance the final results must be
independent of $G$.

The index $\nu$ stands for all solutions included the discrete and
both negative and positive energy solutions of the Dirac equation.  In
Eq.~(\ref{eq_gera}), moreover, $\left\vert i\right\rangle = \left\vert
  n_i \kappa_i m_i \right\rangle$, $\left\vert \nu\right\rangle =
\left\vert n_{\nu} \kappa_{\nu} m_{\nu} \right\rangle$ and $\left\vert
  f\right\rangle = \left\vert n_f \kappa_f m_f \right\rangle$ are the
well--known solutions of the Dirac Hamiltonian for a single electron,
where $n$ and $m$ stand for the principal quantum number and the
one-electron angular momentum projection, respectively. The Dirac
quantum number $\kappa$ is defined by
\begin{equation}
\kappa= \left\{\begin{array}{lcl} \ell \quad&\mbox{if }&j=\ell-1/2\\
                              -(\ell +1)\quad&\mbox{if
                              }&j=\ell +1/2\end{array}\right. \ , 
\label{pkappa}
\end{equation}
where $\ell$ and $j$ are the electron orbital and total angular
momenta, respectively.

If the energy of an intermediate state $E_{\nu}$ is equal to the
energy ($E_{i}-\omega _{1,2}$) in the denominators of Eq.
(\ref{eq_gera}), the differential emission rate has a pole or a
resonant behavior at $E_{\nu}$.  Physically, this occurs when an
intermediate virtual state, between the initial and final states,
coincides with a real state so that the two-photon transition
coincides with the cascade de-excitation process.

For example, in the 2E1 $3s_{1/2} \to 1s_{1/2}$ transition, the shape
of the frequency distribution presents narrow resonances at energies
corresponding to the $3s_{1/2} \to 2p_{1/2,3/2}\to 1s_{1/2}$ cascade.
This effect has been confirmed both experimentally \cite{995} and
theoretically \cite{1952}.

The divergent behavior of the resonant denominator in Eq.
(\ref{eq_gera}) is related to the Green function used in that
expression, which does not take into account the interaction between
the electron and the vacuum fluctuations of the electromagnetic field.
The LPA allows to derive the following expression for the differential
emission \cite{1434,2296}, that takes partially into account this
contribution,
 \begin{eqnarray}
\frac{dw^{\mathrm{LPA}}}{d\omega _{1}}&=& \frac{\omega _{1}\omega _{2}}{(2\pi )^{3}c^{2}}
\left\vert \sum_{\nu} \left( \frac{\left\langle f\left\vert A_{2}^{\ast }\right\vert
\nu\right\rangle \left\langle \nu\left\vert A_{1}^{\ast }\right\vert
i\right\rangle }{V_{\nu}-V_{i}+\omega_{1}}\right. \right. \nonumber \\
 &+&\left.\left. \frac{\left\langle f\left\vert A_{1}^{\ast}\right\vert
\nu\right\rangle \left\langle \nu\left\vert A_{2}^{\ast }\right\vert
i\right\rangle }{V_{\nu}-V_{i}+\omega _{2}}\right)\right|^{2}d\Omega
_{1}d\Omega _{2} ,  
\label{eq_gera_r}
\end{eqnarray}
where 
\begin{equation}
V_{\nu}=E_{\nu}+ \eta_{\nu}\left\{\left\langle
    \nu\left|\sum\mbox{}_{e}\right|\nu\right\rangle+\left\langle
    \nu\left|\prod\mbox{}_{e}\right|\nu\right\rangle\right\} 
% +\left\{\left\langle n\left|\prod \mbox{}_{e}\right|n\right\rangle\right\}  
\label{radia_corr} 
 \end{equation}
with
\begin{equation}
\eta _{\nu}=\left\{ 
\begin{array}{l}
1\mbox{\ \ \ \ if }\nu\mbox{ is a resonant intermediate state} \\ 
0\mbox{\ \ \ \    } \mbox{otherwise } 
\end{array}%
\right. 
,
\end{equation}
and $\left\langle \nu\left|\sum_{e}\right|\nu\right\rangle$ and
$\left\langle \nu\left|\prod_{e}\right|\nu\right\rangle$ are the electron
mean value of the self-energy and vacuum polarization operators in
lowest order for the state $\nu$, respectively.
Both the mean value of the self--energy and vacuum polarization
operators have a real part, $\Delta E_{\nu}$, that is a correction to
the energy $E_{\nu}$. On the other hand, only the self--energy
operator has an imaginary part, $\Gamma_{\nu}/2$, which is the width
of the state $\nu$.

The average decay rate, i.e., the decay rate summed over the final
$m_f$ and averaged over initial $m_i$ ion magnetic sublevels, can be
obtained from Eq.  (\ref{eq_gera_r}) as

\begin{equation}
\frac{d{W}^{\mathrm {LPA}}}{d\omega _{1}} =
         \sum_{L_{1},\lambda _{1},L_{2},\lambda _{2}}
         \frac{d\overline{W}^{\mathrm{LPA}}_{L_{1},\lambda
         _{1},L_{2},\lambda _{2}}}{d\omega _{1}},  
 \label{dW/dw1_final_e}
\end{equation}
where the partial decay rates describing the two-photon transitions
of a given type ($\lambda$) and multipolarity ($L$) are given by
\begin{eqnarray}
\frac{d\overline{W}^{\mathrm {LPA}}_{L_{1},\lambda _{1},L_{2},\lambda
    _{2}}}{d\omega _{1}} &=&
\frac{\omega _{1}\omega _{2}}{(2\pi
)^{3}c^{2}(2j_{i}+1)}  \nonumber \\
&& \times \sum_{j_{\nu}}\left[\left\vert
\overline{S}^{j_{\nu}}(2,1)\right\vert ^{2}+\left\vert
    \overline{S}^{j_{\nu}}(1,2)\right\vert ^{2}+ \right. \nonumber \\ 
&&+2\sum_{j_{\nu}^{\prime }}d(j_{\nu},j_{\nu}^{\prime }) \nonumber \\
&& \times \left\{ \mbox{Re}\left[\overline{S}^{j_{\nu}}(2,1)\right]  \mbox{Re}\left[
      \overline{S}^{j_{\nu}^{\prime}}(1,2)\right] \right. \nonumber\\
&&\left.+\left. \mbox{Im}\left[\overline{S}^{j_{\nu}}(2,1)\right]  \mbox{Im}\left[
    \overline{S}^{j_{\nu}^{\prime }}(1,2)\right]
    \right\} \right]. \nonumber\\
\label{SJ_com}
\end{eqnarray}
Here we define
\begin{equation}
d(j,j^{\prime })=(-1)^{2j^{\prime }+L_{1}+L_{2}}\left[ j,j^{\prime }\right]
^{1/2}\left\{ 
\begin{array}{ccc}
j_{f} & j^{\prime } & L_{1} \\ 
j_{i} & j & L_{2}%
\end{array}%
\right\} ,
\label{d(j,j)}
\end{equation}
which represents the angular coupling, and
\begin{eqnarray}
\overline{S}^{j}(2,1) &=&\sum_{n_{\ell}}\frac{\overline{M}_{f,n_{\ell}}^{(\lambda
_{2},L_{2})}(\omega_{2}) \ \overline{M}_{n_{\ell},i}^{(\lambda
_{1},L_{1})}(\omega_{1})}{V_{n_{\ell}}-V_{i}+\omega
_{1}}  \nonumber \\ 
%&& \times 
%\Delta ^{j}(2,1)\Pi ^{\ell}(2,1).  
%\label{SjN_r}
%\end{eqnarray}
%
%\begin{eqnarray}
%\Delta ^{j}(2,1) &=&
&& \times \frac{4\pi \lbrack j_{i},j,j_{f}]^{1/2}}{%
[L_{1},L_{2}]^{1/2}} \pi _{i}^{\ell}(1)\pi _{f}^{\ell}(2)  \nonumber \\
& & \times 
\left( 
\begin{array}{ccc}
j_{f} & L_{2} & j \\ 
\frac{1}{2} & 0 & - \frac{1}{2} %
\end{array}%
\right) 
\left( 
\begin{array}{ccc}
j & L_{1} & j_{i} \\ 
\frac{1}{2} &  0 & - \frac{1}{2} 
\end{array}%
\right) ,  \nonumber \\
%&& \times 
%\pi _{i}^{\ell}(1)\pi _{f}^{\ell}(2),
%
\label{SjN_r}
\end{eqnarray}
with
\begin{equation}
\pi^{\ell}_{k}(t)= \left\{\begin{array}{c} 1\quad\mbox{if
                              }\ell_{k}+\ell+L_{t}+\lambda_{t}=\mbox{odd} \\
                             0\quad\mbox{if
                              }\ell_{k}+\ell+L_{t}+\lambda_{t}=\mbox{even}\end{array}\right.
                              .
\label{parity2}
\end{equation}
$\overline{S}^{j}(1,2)$ is analogously defined. The notation
$[j,k,..]$ means $(2j +1)(2k+1) \ldots$, $(\cdots)$ are the $3j$ symbols
and $\left\{ \cdots \right\}$ the $6j$ symbols.

The radial matrix elements $\overline{M}_{f,i}^{(\lambda,L)}$ in Eq.
(\ref{SjN_r}) are defined by
\begin{eqnarray}
\overline{M}_{f,i}^{(1,L)} &=&\left( \frac{L}{L+1}\right) ^{1/2}\left[
\left( \kappa_{f}-\kappa_{i}\right) I_{L+1}^{+}+\left( L+1\right) I_{L+1}^{-}%
\right]  \nonumber \\
&&-\left( \frac{L+1}{L}\right) ^{1/2}\left[ \left( \kappa_{f}-\kappa_{i}\right)
I_{L-1}^{+}-LI_{L-1}^{-}\right],\nonumber \\
\label{M_al_bet1}
\end{eqnarray}
\begin{equation}
\overline{M}_{f,i}^{(0,L)}=\frac{2L+1}{\left[ L\left( L+1\right) \right]
^{1/2}}\left( \kappa_{f}+\kappa_{i}\right) I_{L}^{+} , 
\label{M_al_bet}
\end{equation}
and
\begin{eqnarray}
\overline{M}_{f,i}^{(-1,L)} &=&G[(2L+1)J^{\left( L\right) } \nonumber \\
&&+\left( \kappa_{f}-\kappa_{i}\right) \left( I_{L+1}^{+}+I_{L-1}^{+}\right)
\nonumber \\
&&-LI_{L-1}^{-}+\left( L+1\right) I_{L+1}^{-}].  
\label{M_min} 
\end{eqnarray}
$L$ is the photon angular momentum and $\lambda$ stands for the
electric ($\lambda$=1), magnetic ($\lambda$=0) and the longitudinal
($\lambda=-1$) terms. We used the notation given by Rosner and Bhalla
\cite{17} for the integrals the $I_{L}^{\pm }\left( \omega \right) $
and $J_{L}\left( \omega \right) $.  The parity selection rules
(\ref{parity2}) follow from the calculation of the reduced matrix
elements expressed in Eq. (\ref{eq_gera}).

We emphasize that the term $\pi _{i}^{\ell}(1)\pi _{f}^{\ell}(2)$ in
(\ref{SjN_r}) is not given explicitly in the Goldman and Drake article
\cite{3}, which could lead to some ambiguity in the choice of the
intermediate states for the evaluation of the $\overline{S}^{j}(2,1)$
and $\overline{S}^{j}(1,2)$ terms in a generic transition.

Usually, it is convenient to express the results in terms of the
electric ($E$) and magnetic ($M$) multipole contributions. The total
decay rate (integrated over the photon energy) for a transition in
which one photon $\Theta _{1}L_{1}$ and one photon $\Theta _{2}L_{2}$
are emitted, where $\Theta _{i}=E,M$ stands for the electric and
magnetic multipole type, respectively, is given by
\begin{eqnarray}
%\begin{equation}
\overline{W}^{\mathrm {LPA}}_{\Theta _{1}L_{1}\Theta _{2}L_{2}} &=&
\sum_{\lambda _{\Theta _{1}},\lambda _{\Theta _{2}}}
\overline{W}^{\mathrm {LPA}}_{L_{1},\lambda _{\Theta
    _{1}},L_{2},\lambda _{\Theta _{2}}} \nonumber\\
&=&\sum_{\lambda _{\Theta _{1}},\lambda _{\Theta _{2}}}
\int_{0}^{\omega _{t}} 
\frac{d\overline{W}^{\mathrm {LPA}}_{L_{1},\lambda _{\Theta _{1}},L_{2}\lambda _{\Theta
   _{2}}}} {d\omega _{1}} d\omega _{1}, \nonumber \\ 
\label{rrii6}
\end{eqnarray}
with
\begin{equation}
\left\{ 
\begin{array}{lcl}
\lambda _{\Theta _{i}}=-1,1 & \mbox{\ \ \ if\ \ \ } & \Theta_{i}=E  \\ 
\lambda _{\Theta _{i}}=0 & \mbox{\ \ \ if\ \ \ } & \Theta _{i}=M
\end{array}
\right.  
,
\label{E_M}
\end{equation}
and $\omega _{t}$ is the energy of the two-photon transition, which is
given, in a.u., by
\begin{equation}
\omega_{t}= \omega_{1}+ \omega_{2}=E_{f}-E_{i}
\label{w_t}
\end{equation}
using Eq. (\ref{con_ener}).

Finally, the total spontaneous emission probability per unit time
for a two-photon transition is obtained by summing over all allowed
multipole components,

\begin{equation}
W^{\mathrm {LPA}}=\sum_{\mbox{all \ }\Theta _{1}L_{1},\Theta _{2}L_{2}}
t_{\Theta _{1}L_{1},\Theta _{2}L_{2}}
\overline{W}^{{\mathrm {LPA}}}_{ \Theta _{1}L_{1}\Theta _{2}L_{2}},
\label{tput}
\end{equation}
where
\begin{equation}
t_{\Theta_{1}L_{1},\Theta _{2}L_{2}}=\left\{ 
\begin{array}{c}
1\mbox{ \ \ \ \ \ \ if \ }\Theta _{1}L_{1}\neq \Theta _{2}L_{2}\mbox{\ } \\ 
1/2\mbox{\ \ \ \ if \ }\Theta _{1}L_{1}=\Theta _{2}L_{2}\mbox{ }%
\end{array}
\right.  
.
\label{twice}
\end{equation}
The factor $1/2$ is included to avoid counting twice each pair, when
both photons have the same characteristics.

Another method for dealing with resonances was developed by Jentschura
and co-workers \cite{2285,2491} using a procedure based on two--loop
self--energy (TLA). They obtained an expression similar to Eq.
(\ref{eq_gera}) for evaluating a nonresonant component of the
two--photon decay rate, given by
\begin{equation}
w^{\textrm{TLA}} =\lim_{\epsilon\rightarrow 0}
\mbox{Re}\int_0^{\omega_{t}}d\omega _{1} \frac{\omega _{1}\omega
  _{2}}{(2\pi )^{3}c^{2}} \mathcal{S}_{if}d\Omega _{1}d\Omega_{2} ,  
\label{eq_gera_Jun}
\end{equation}% 
The function $\mathcal{S}_{if}$ is given, as in Ref. \cite{2491}, by
\begin{eqnarray}
\mathcal{S}_{if} &=&  \left( \sum_{\nu} \left\{\frac{\left\langle
        f\left|A_{2}^{*}\right| \nu \right\rangle
        \left\langle \nu \left|A_{1}^{*}\right|i \right\rangle}
        {E_{\nu}-E_{i}+\omega_{1}-i\epsilon}
        \right. \right.  \nonumber \\
        &+& 
        \left. \left. \frac{\left\langle 
        f\left|A_{1}^{*}\right|\nu \right\rangle\left\langle
        \nu \left|A_{2}^{*}\right|i\right\rangle}{E_{\nu}-E_{i}+\omega_{2}-i\epsilon}
        \right\}\right)^2  .
\label{eq_gera_Sif}
\end{eqnarray}

Using this approach one obtains finite results since the integration
over the frequency $\omega_{1}$ is displaced by a infinitesimal
quantity, $\epsilon$, from the resonance poles, provided the limit is
not permuted with the integration.

If one considers a nonresonant transition such as $2s_{1/2}\rightarrow1s_{1/2}$,
then the limit can be permuted with the integration and
Eq.~(\ref{eq_gera_Jun}) reduces to Eq.~(\ref{eq_gera}), and both
approaches gives the same result.

%
%---------------------------------------------------------------------
%
\subsubsection{\label{Integration method}Integration method for
  resonant intermediate states} 
%\label{Integration method}
%
For resonant transitions, Eq.~(\ref{SJ_com}) produces sharp peaks near
the resonant frequencies, which requires special attention in the
integration over the photon energy $\omega_1$ in Eq. (\ref{rrii6}) to
avoid meaningless results for the total decay rate. Near a resonant
frequency $\omega _{\textrm{R}}^{j_{\nu}}$, Eq.~(\ref{SJ_com}) can be
written as

\begin{eqnarray}
\frac{d\overline{W}^{\textrm{LPA}}_{L_{1},\lambda _{1},L_{2},\lambda _{2}}}{d\omega _{1}}%
&=&\sum_{j_{\nu}}g^{j_{\nu}}\left( \omega _{1}\right) \nonumber \\
 &=&\sum_{j_{\nu}}\frac{f^{j}\left( \omega
_{1}\right) }{\left( \omega _{1}-\omega _{\textrm{R}}^{j_{\nu}}\right) ^{2}+\left( \frac{%
\Gamma _{\textrm{R}}^{j_{\nu}}}{2}\right) ^{2}},
\label{smooth_isola}
\end{eqnarray}
where $f^{j}\left( \omega _{1}\right) $ is a smooth function; the
resonant behavior is given by the denominator. Consequently, the
function $f^{j}\left( \omega _{1}\right) $ can be expanded in a Taylor
series around the resonant frequency $\omega _{\textrm{R}}^{j_{\nu}}$.
Notice that the shape in the right-hand side of
Eq.~(\ref{smooth_isola}) is not a Lorentz profile since
$f^{j}(\omega_{1})$ depends on $\omega_{1}$ and so, the peak profile
is asymmetric.  Subtracting the first two terms of the expansion on
$g^{j}\left( \omega _{1}\right) $ we obtain a smooth function,
$h^{\textrm{LPA}}\left( \omega _{1}\right)$, which does not contain a
resonant behavior. It is defined as
\begin{eqnarray}
h^{\textrm{LPA}}\left( \omega _{1}\right) &=&\sum_{j_{\nu}} \left[g^{j_{\nu}}\left( \omega _{1}\right)
-\frac{a_{0}^{j_{\nu}}}{\left( \omega 
_{1}-\omega _{\textrm{R}}^{j_{\nu}}\right) ^{2}+\left( \frac{\Gamma
  _{\textrm{R}}^{j_{\nu}}}{2}\right) ^{2}} \right. \nonumber \\  
&-& \left. \frac{a_{1}^{j_{\nu}}\left( \omega _{1}-\omega _{\textrm{R}}^{j_{\nu}}\right) }{\left( \omega
_{1}-\omega _{\textrm{R}}^{j_{\nu}}\right) ^{2}+\left( \frac{\Gamma
  _{\textrm{R}}^{j_{\nu}}}{2}\right) ^{2}} \right].
\label{fun_h}
\end{eqnarray}
The coefficients $a_{0}^{j}$ and $a_{1}^{j}$ are derived from the Taylor
expansion of $f^{j}\left( \omega _{1}\right) $ around $\omega
_{\textrm{R}}$:
\begin{eqnarray}
a_{0}^{j} &=&f^{j}\left( \omega _{\textrm{R}}\right) =g^{j}\left( \omega _{\textrm{R}}\right)
 \left( \frac{\Gamma _{\textrm{R}}^{j}}{2}\right) ^{2},\nonumber  \\
a_{1}^{j} &=&\left\{ \frac{d}{d\omega _{1}}f^{j}\left( \omega _{1}\right)
\right\} _{\omega _{\textrm{R}}} \nonumber \\ 
&=&\left\{ \frac{d}{d\omega _{1}}g^{j}\left( \omega
_{1}\right) \right\} _{\omega _{\textrm{R}}} \left( \frac{\Gamma _{\textrm{R}}^{j}}{2}
\right) ^{2}.
\label{aaa01}
\end{eqnarray}
The expressions of the derivatives of the matrix elements used to
evaluate $a_{1}^{j}$ are presented in the Appendix.

To obtain the decay rate $\overline{W}^{\textrm{LPA}}_{L_{1},\lambda
  _{1},L_{2},\lambda _{2}}$ we must add to the integral of the smooth
function $h^{\textrm{LPA}}$ the two terms
$\mathfrak{h}^{\textrm{LPA}}_0$ and $\mathfrak{h}^{\textrm{LPA}}_1$
evaluated analytically, i.e.,
\begin{equation}
\overline{W}^{\textrm{LPA}}_{L_{1},\lambda _{1},L_{2},\lambda _{2}} =
\mathfrak{h}^{\textrm{LPA}} +
\mathfrak{h}^{\textrm{LPA}}_{0}+\mathfrak{h}^{\textrm{LPA}}_{1}, 
\label{sum_h}
\end{equation}
where
\begin{equation}
\mathfrak{h}^{\textrm{LPA}} = \int_{0}^{\omega _{t}}h^{\textrm{LPA}} \ d\omega _{1}
\label{hh}
\end{equation}

\begin{eqnarray}
\mathfrak{h}^{\textrm{LPA}}_{0} &=&\sum_{j_{\nu}}a_{0}^{j_{\nu}}\int_{0}^{\omega _{t}}\frac{1}{\left|\omega
_{1}-\omega _{\textrm{R}}^{j_{\nu}}-i\frac{\Gamma _{\textrm{R}}^{j_{\nu}}}{2} \right|^{2}}d\omega _{1}  \nonumber \\
&=&\sum_{j}\frac{2a_{0}^{j_{\nu}}}{\Gamma _{\textrm{R}}^{j}}\left[ \arctan \left( \frac{
2\left( \omega _{1}-\omega _{\textrm{R}}^{j_{\nu}}\right) }{\Gamma _{\textrm{R}}^{j_{\nu}}}\right) \right]
_{0}^{\omega _{t}},
\label{h0}
\end{eqnarray}

\begin{eqnarray}
\mathfrak{h}^{\textrm{LPA}}_{1} &=&\sum_{j_{\nu}}a_{1}^{j_{\nu}}\int_{0}^{\omega _{t}}\frac{\left( \omega
_{1}-\omega _{\textrm{R}}^{j_{\nu}}\right) }{\left|\omega _{1}-\omega _{\textrm{R}}^{j_{\nu}}-i\frac{
\Gamma _{\textrm{R}}^{j_{\nu}}}{2}\right|^{2}}d\omega _{1}  \nonumber \\
&=&\sum_{j}\frac{a_{1}^{j_{\nu}}}{2}\ln \left[ \frac{\left( \omega _{t}-\omega
_{\textrm{R}}^{j_{\nu}}\right) ^{2}+\left( \Gamma _{\textrm{R}}^{j_{\nu}}/2\right) ^{2}}{\left( \omega
_{\textrm{R}}^{j_{\nu}}\right) ^{2}+\left( \Gamma _{\textrm{R}}^{j_{\nu}}/2\right) ^{2}}\right]. 
\label{h111}
\end{eqnarray}

We note that $a_{0}^{j}$ is given approximately (unless we consider
the limit $\Gamma\rightarrow0$, in which case it is given exactly) by
\begin{equation}
a_{0}^{j}\approx\frac{w^{\lambda_{2},L_{2}}_{i\rightarrow
    r}w^{\lambda_{1},L_{1}}_{r\rightarrow f}}{2\pi }, 
\label{a0_aprox}
\end{equation}
where the term $w^{\lambda_{k},L_{k}}_{i\rightarrow f}(\omega)$ is the
decay rate from a initial to a final state through the emission of one
photon, which is given, in a.u., by \cite{6}
\begin{equation}
w^{\lambda L}_{i\rightarrow f}(\omega)=
   \frac{2\omega \lbrack j_{f}]}
   {c[L]}\left( 
\begin{array}{ccc}
j_{f} & L & j_{i} \\ 
\frac{1}{2} & 0 & -\frac{1}{2}%
\end{array}%
\right) ^{2}
   \left\vert \overline{M}^{\lambda L}_{f,i}\right\vert ^{2}.
\label{one_pho}
\end{equation}
Using this result we can write the term
\begin{equation}
\frac{d\mathfrak{h}^{\textrm{LPA}}_{0}}{d\omega_{1}}
\approx\sum_{j_{\nu}}\frac{1}{2\pi}\frac{w_{i\rightarrow r}w_{r\rightarrow
    f}}{\left(\omega 
_{1}-\omega _{\textrm{R}}^{j_{\nu}}\right)^{2}+ \left(\frac{\Gamma
  _{\textrm{R}}^{j_{\nu}}}{2}\right) ^{2}},  
\label{h0_cas}
\end{equation}
and identify $\mathfrak{h}^{\textrm{LPA}}_{0}$ as a cascade transition
rate contributions.

Applying a similar approach to $w^{\textrm{TLA}}$ given by Eq.
(\ref{eq_gera_Jun}) we obtain a smooth function $h^\textrm{TLA}$ as in
the LPA. One difference between the two approaches is in the term of
order $\sim \left(\Gamma_{\textrm{R}}/2\right)^2$, which appears in
the denominator of Eq.~(\ref{fun_h}) and results from considering the
infinitesimal quantity $\epsilon$ finite, i.e., taking the role of a
level width ($\epsilon\rightarrow\Gamma$). In the present evaluation
we obtain the function $h^{\textrm{TLA}}$ by replacing
$\Gamma_{\textrm{R}}\rightarrow q\Gamma_{\textrm{R}}$, where $q$ is a
parameter that can be made arbitrarily small. We thus obtain
convergence since the difference in $h^{\textrm{TLA}}$ using $q=1$ or
$q=10^{-2}$ is in the fifth digit.  For $q=10^{-2}$ and $q=10^{-3}$
the difference in $h^{\textrm{TLA}}$ is in the ninth digit. So we
conclude that using $h^{\textrm{LPA}}$ defined in Eq.  (\ref{fun_h})
with $q=10^{-2}$, is a good approximation for the function
$h^{\textrm{TLA}}$. Another difference between TLA and LPA is the
inclusion of radiative corrections Re [SE] and VP, which for values of
$Z$ as high as $92$ changes the value of $h$ from one approach to
another in the second digit.
The major difference between the two approaches is in the integral
$\mathfrak{h}_{0}$, which in TLA is given by
\begin{equation}
\mathfrak{h}^{\textrm{TLA}}_{0} =\sum_{j_{\nu}}a_{0}^{j_{\nu}}\frac{1}{\omega
    _{\textrm{R}}^{j_{\nu}}\left( \omega  _{\textrm{R}}^{j_{\nu}}-\omega _{t}\right) }, 
\end{equation}
which comes from the different ways the pole regularization is done.

Notice that the terms $\mathfrak{h}_{0}^{\textrm{LPA}}$ and
$\mathfrak{h}_{0}^{\textrm{TLA}}$ are related by
\begin{equation}
\mathfrak{h}_{0}^{\textrm{LPA}}=\mathfrak{h}_{0}^{\textrm{TLA}}+\sum_{j_{\nu}}\frac{2\pi
  a_{0}^{j_{\nu}}}{\Gamma _{\textrm{R}}^{j_{\nu}}} + O\left( \Gamma _{\textrm{R}}^{j_{\nu}}\right) ,
\label{h_LPA_JA}
\end{equation}
which shows that the difference between
$\mathfrak{h}_{0}^{\textrm{LPA}}$ and
$\mathfrak{h}_{0}^{\textrm{TLA}}$ is mainly due to the second term on
the left side of Eq.~(\ref{h_LPA_JA}) (since $\Gamma _{\textrm{R}}^{j}
\ll 1$) or by the product of one-photon transitions (cascade
process).

On the other hand, the integral $\mathfrak{h}_{1}$ is given in the TLA
approach by
 \begin{equation}
\mathfrak{h}^{\textrm{TLA}}_{1} =\sum_{j_{\nu}}a_{1}^{j_{\nu}}\ln \left[ \frac{ \omega _{t}-\omega
_{\textrm{R}}^{j_{\nu}}}{ \omega_{\textrm{R}}^{j}}\right].
\label{h1_ja}
\end{equation}
This expression can be obtained from Eq.~(\ref{h111}) by taking
$\Gamma_{\textrm{R}}\rightarrow0$.

Considering that
\begin{equation}
\mathfrak{h}^{\textrm{TLA}} = \int_{0}^{\omega _{t}}h^{\textrm{TLA}}\ d\omega _{1}
\label{hhja}
\end{equation}
the decay rate in the TLA, $\overline{W}^{\textrm{TLA}}_{L_{1},\lambda
  _{1},L_{2},\lambda _{2}}$, is given by an expression similar to Eq.
(\ref{sum_h}), in which the LPA contributions are replaced by the
correspondent TLA contributions.
  
As we will see in Section \ref{Numer_result}, these differences on the
sum of $\mathfrak{h}$ and $\mathfrak{h1}$ do not carry any sizable
difference between the LPA and TLA methods for low-$Z$ ions, but lead to
slight discrepancy for heavier systems.

%
%%%%%%%%%%%%%%%%%%%%%%%%%%%%%%%%%%%%%%%%%%%%%%%%%%%%%%%%%%%%%%%%%%%%%%%%%%%%%%%%%%%%%%%%%%%%%
%---------------------------------------------------------------------
%
\subsection{\label{sec_B_splines}Solution of the Dirac-Fock equation
  on a B-splines basis set} 
%\label{sec_B_splines}

To make the numerical evaluation we consider that the atom, or ion, is
enclosed in a finite cavity with a radius large enough to get a good
approximation of the wavefunctions, with some suitable set of boundary
conditions, which allows for discretization of the continua.

Let us denote by $\left\{ \phi _{n}^{i}(r),\mbox{ }i=1,.., 2N\right\}$
a set of solutions of the Dirac-Fock equation, where $n$ is the level,
$i$ the position of the solution in the set and $N$ the number of
functions in the basis set. 

For each $n$ value the set is complete,
and $\phi _{n}^{i}(r)$ obeys the equation \cite{47}
\begin{equation}
\left[ 
\begin{array}{cc}
\frac{V(r)}{c} & \frac{d}{dr}-\frac{\kappa}{\textrm{R}} \\ 
\\
-\left( \frac{d}{dr}+\frac{\kappa}{\textrm{R}} \right) & -2c+\frac{V(r)}{c}
\end{array}
\right] \phi _{n}^{i}(r)=\dfrac{ \varepsilon _{n}^{i}}{c} \phi _{n}^{i}(r) ,
\label{eq_dirac}
\end{equation}
where the energy $E_{n}^{i}$ was replaced by $\epsilon_{n}^{i}
=E_{n}^{i}-mc^{2}$.  
The potential $V(r)$ is given by a Coulomb potential assuming a
uniform nuclear charge distribution for a finite nucleus and $\kappa$
is given by Eq. (\ref{pkappa}). A complete set spans both positive and
negative solutions. Solutions labeled by $i=1,...,N$ describe the
continuum $\varepsilon _{n}^{i}<-2mc^{2}$ and solutions labeled by
$i=N+1$,$..., 2N$ describe bound states (the few first ones) and the
continuum $\varepsilon _{n}^{i}>0$.
%using the conventions of the
%equation (\ref{eq_dirac}). 
%
For practical reasons, such as easy numerical implementation, this set
of solutions is itself expressed as linear combination of another
basis set. We have chosen the B-splines basis set and we used the
derivation of the solution of equation (\ref{eq_dirac}) in terms of
B-splines described by Johnson, Blundell and Sapirstein
\cite{8}.
%
%=====================================================================
%
\section{\label{Numer_result}Numerical results and discussion}
%\label{Numer_result}
%
By taking $\eta_{\nu}=0$ in Eq.  (\ref{radia_corr}) we may calculate
the two-photon decay rates without accounting for radiative
corrections. In this case, we have verified with respect to variations
of the gauge parameter ($G=0$ for velocity gauge and $G=\sqrt{2}$ for
length gauge), the radius of the cavity ($R$), and the basis set
parameters (the number $(ns$) and the degree ($k$) of the B-splines),
the stability and accuracy to six digits on the calculation of
Eq.~(\ref{SJ_com}) for a nonresonant states and for a frequency
$\omega _{1}$.

The parameters used in the calculation of the results presented in
this work are $k=9$, $ns=60$ and $R=60$ a.u.. 
The integration over the photon frequency has been performed using a 15
points Gauss-Legendre algorithm for the nonresonant transitions.
%
%
%---------------------------------------------------------------------
%s

\subsection{Nonresonant transitions}

For the nonresonant $2s_{1/2}\rightarrow1s_{1/2}$ and
$2p_{1/2}\rightarrow1s_{1/2}$ transitions, we use Eq.
(\ref{dW/dw1_final_e}) for both the decay rate values for various
multipole combinations $(\Theta_1 L_1, \Theta_2 L_2)$ and the
frequency distribution.

The most significant multipole combinations included in the
calculation of the two-photon decay rates of the $2p_{1/2}\rightarrow
1s_{1/2}$ transition are presented in Table \ref{tab_2p}. The
magnitude of the multipole combinations not listed in this table are,
at least, four orders of magnitude smaller than the most significant

The values given by Labzowsky {\it et al.} in \cite{1418} were
obtained using expressions similar to the ones used by Goldman and
Drake \cite{3}. In Ref.  \cite{1907}, the results were obtained using
nonrelativistic Coulomb Green's function, which, for high values of
$Z$ such as 92 leads to inaccurate values. The relative difference
between our results and the results in Ref. \cite{1418} is in the
range of $0.1-0.4\%$. We observe that, for the three studied $Z$
values, more than $99\%$ of the total decay rate is due to the
multipole contributions $E1M1$ (about $60\%$) and $E1E2$ (about
$40\%$).
The fact that the two multipole combinations $E1M1$ and $E1E2$ give
almost the same contribution is somewhat expected since $M1$ and $E2$
have the same order of magnitude in the decomposition of the photon
field \cite{12}.
On the other hand, a comparison between the listed most significant
($E1M1$) and less ($E2E3$) significant contributions reveals that the
relative importance of the latter increases with $Z$, being 12, 5 and
4 orders of magnitude smaller than the former for $Z=1$, $Z=40$, and
$Z=92$, respectively.

In Table \ref{Tab_23_n2}, we report the two-photon total decay rates
for $2p_{1/2}\rightarrow 1s_{1/2}$ and $2s_{1/2}\rightarrow 1s_{1/2}$
transitions.  Enough multipoles have been included in the calculation
of the total 2-photon decay rates to reach an accuracy of six digits.
The values for the $2s_{1/2}\rightarrow1s_{1/2}$ transition differ
slightly from the ones in our previous work \cite{388} due to use of
the most recent values of physical constants \cite{1183}, such as the
fine-structure constant.

It sould be mentioned that the interest in the transition
$2p_{1/2}\rightarrow1s_{1/2}$, and other two photon forbided
transitions, is only academic since the transition is suppressed by
selection rules and this channel is in direct competition with an
allowed one-photon transition.

To present the spectral (or frequency) distribution for a specific
value of $Z$ is convenient to express the results in $\psi(y,Z)$ as
suggested by Spitzer and Greenstein \cite{159}
\begin{equation}
\frac{dW}{dy}=\left( \frac{9}{2^{10}}\right) \left( Z\alpha \right)
^{n}\psi (y,Z),
%\quad \mbox{Ry} ,  
\label{esp}
\end{equation}
where $y=\omega /\omega _{fi}$ is the fraction of the photon energy
carried by one of the photons and $\omega _{fi}$ is the energy of the
transition.  In case of an even$\rightarrow $even (or odd$\rightarrow
$odd) transition, the major multipole contribution $2E1$ scales as
$Z^{6}$ and, consequently, $n=6$. For a even$\rightarrow $ odd (or
odd$\rightarrow $even) transition, both $E1M1$ and $E1E2$ scale as
$Z^{8}$.

\begin{figure}[htb]
\centering
\includegraphics[clip=true,width=15cm]{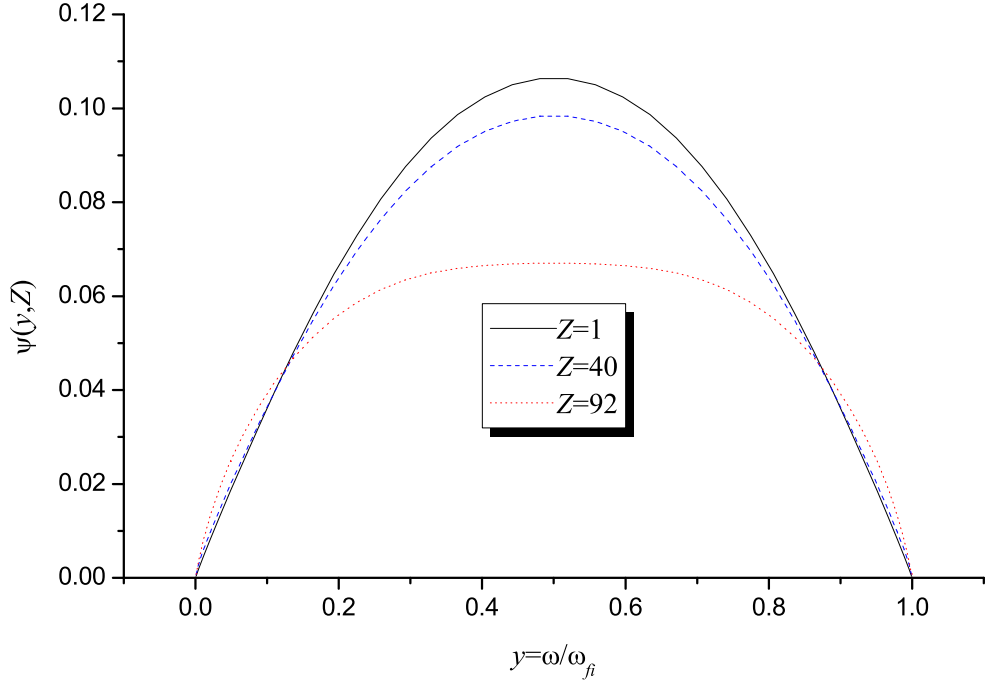}
\caption{Spectral distribution function $\psi (y,Z)$, defined by
  Eq. (\ref{esp}), for the transition $2p_{1/2} \rightarrow 1s_{1/2}$ at
  $Z$=1, 40, and 92. The variable $y=\omega/\omega_{fi}$ is the
  fraction of the photon energy carried by one of the two-photons.}
%Same as Fig. \ref{fig_2s} for the transition $2p_{1/2}
%  \rightarrow 1s_{1/2}$. }
\label{fig_2p_1s}
\end{figure}

In Fig. \ref{fig_2p}, the frequency distribution of the multipole
contributions $E1M1$, $M1E1$, $E1E2$ and $E2E1$ for the transition
$2p_{1/2}\rightarrow 1s_{1/2}$ are presented. Although each one of
these four most significant contributions is asymmetric, the sum of
each pair ($E1M1$, $M1E1$) and ($E1E2$, $E2E1$) is symmetric around
$y=0.5$.  Therefore, the total frequency distribution is also
symmetric around the $y=0.5$ value, as can be seen in Fig.
\ref{fig_2p_1s}, in which we also notice the $Z$ dependence of the
shape predicted by Goldman and Drake \cite{3} for the
$2s_{1/2}-1s_{1/2}$ transition.
%
%
%---------------------------------------------------------------------
%

\subsection{Resonant transitions}

\begin{figure}[htb]
  \centering
\includegraphics[clip=true,width=15cm]{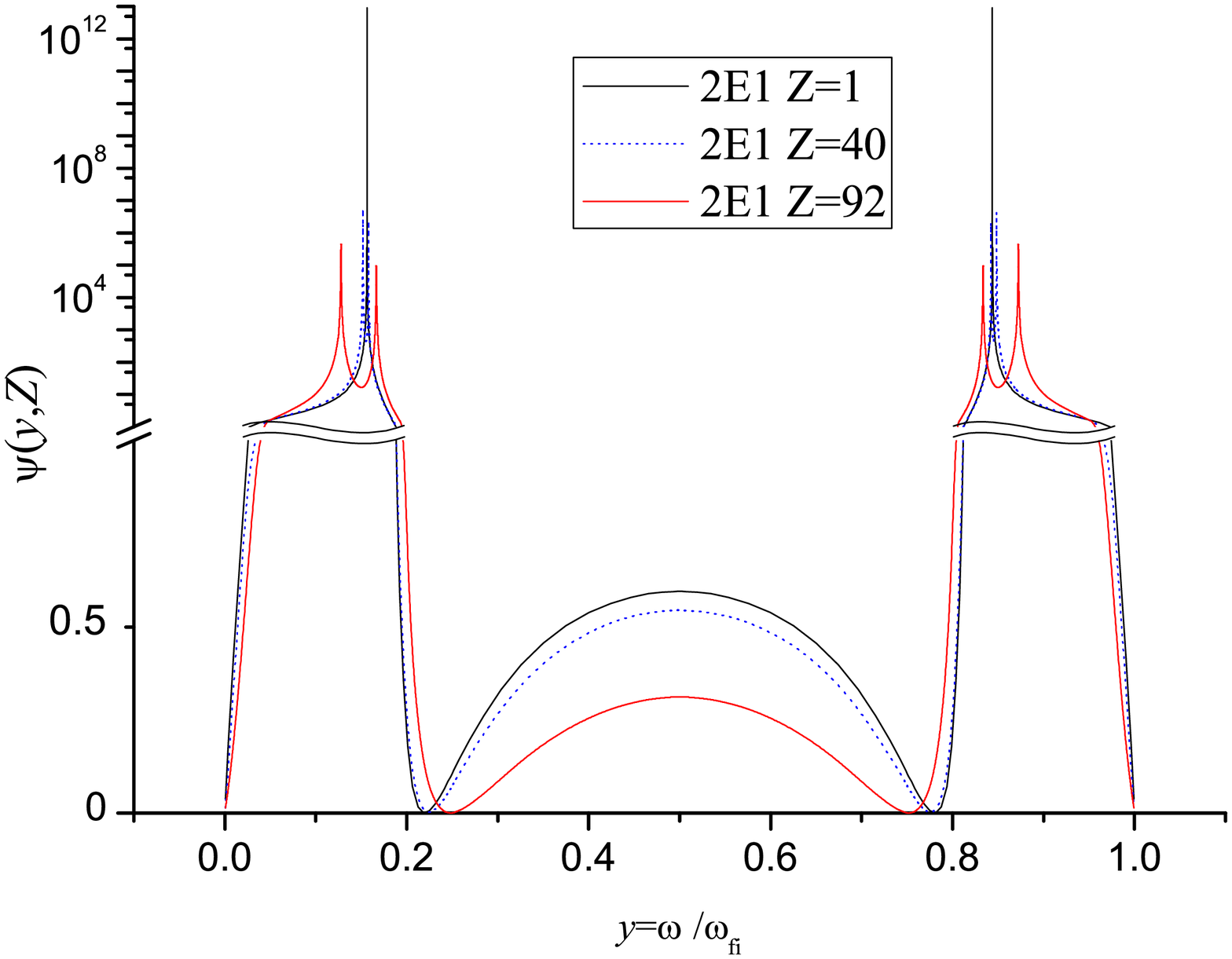}
\caption{
  Spectral distribution function $\psi (y,Z)$, defined by Eq.
  (\ref{esp}), of the $2E1$ contribution for the transition $3s_{1/2}
  \rightarrow 1s_{1/2}$ at $Z$=1, 40, and 92. The variable
  $y=\omega/\omega_{fi}$ is the fraction of the photon energy
  carried by one of the two-photons.}
\label{fig_3s}
\end{figure}

\begin{figure}[htb]
  \centering
\includegraphics[clip=true,width=15cm]{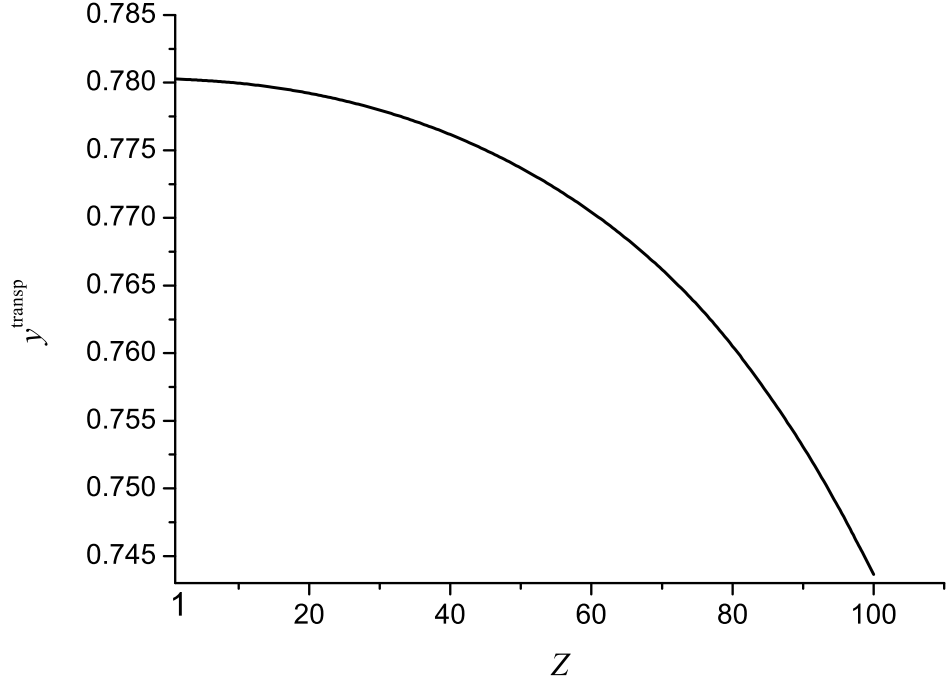}
\caption{Transparency frequency, $y^{\mbox{\small transp}}$, of the transition
  $3s_{1/2}\rightarrow1s_{1/2}$ as function of the atomic number $Z$.}
\label{fig_trans_Z}
\end{figure}

After this brief discussion of the nonresonant $2s_{1/2} \rightarrow
1s_{1/2}$ and $2p_{1/2} \rightarrow 1s_{1/2}$ two-photon transitions,
we now turn to the evaluation of the differential total decay rates
for the higher excited ionic states. In Fig.  \ref{fig_3s}, for
example, we display the spectral distribution for the $3s_{1/2}
\rightarrow 1s_{1/2}$ transition. We notice several features that are
not found in the corresponding plot for the $2s_{1/2} \rightarrow
1s_{1/2}$ transition.  In particular, the $\psi_{3s_{1/2} \rightarrow
  1s_{1/2}} (y,Z)$ function exhibits sharp peaks, which are due to the
$3s_{1/2} \to 2p_{1/2,3/2} \to 1s_{1/2}$ cascade. Furthermore, we
observe that at $Z=92$ each of the two resonances splits in two due to
the spin-orbit interaction and the frequency gap in each pair is
exactly equal to the difference between the states $2p_{3/2}$ and
$2p_{1/2}$, respectively.
In addition, besides the zeroes at the endpoints, there are two more
minima at $y=0.219733$ and $0.780267$. Such minima were observed in
two-photon spectra by Tung \textit{et al} \cite{95,1906}, and they
were referred to as ``transparencies''. In Table \ref{tab_transp} we
list the transparencies for several two-photon transitions obtained in
this work by other authors. Their relative differences are smaller
than 0.01\% for $Z=1$. To the best of our knowledge, there are no
published data for other $Z$ values.
In Fig.~\ref{fig_trans_Z} we plot the transparency frequency,
$y^{\mbox{\small transp}}$, of the transition
$3s_{1/2}\rightarrow1s_{1/2}$ as function of $Z$. We notice that the
transparency values scale with $Z^{2}$ as the transition energy.
  
\begin{figure}[htb]
\centering
\includegraphics[clip=true,width=15cm]{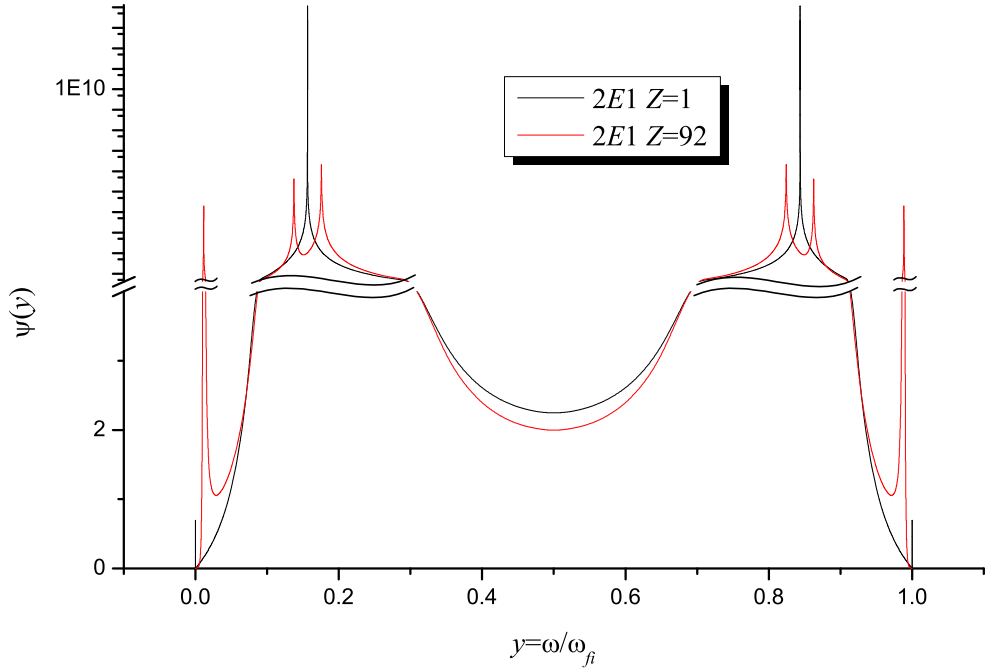}
\caption{Same as Fig. \ref{fig_3s} for the transition $3d_{3/2} \rightarrow 1s_{1/2}$. }
\label{fig_3d_1s}
\end{figure}

In contrast to the $3s_{1/2} \to 1s_{1/2}$, the spectral distribution
for the $3d_{3/2}\rightarrow 1s_{1/2}$ transition, plotted in Fig.
\ref{fig_3d_1s}, exhibits only the resonant behavior as mentioned in
Ref.~\cite{95}, which is due to the fine-structure splitting between
$2p_{1/2}$ and $2p_{3/2}$ and $3p_{1/2}$ and $3d_{3/2}$ states.

\begin{figure}
%[htb]
\centering
\includegraphics[clip=true,width=15cm]{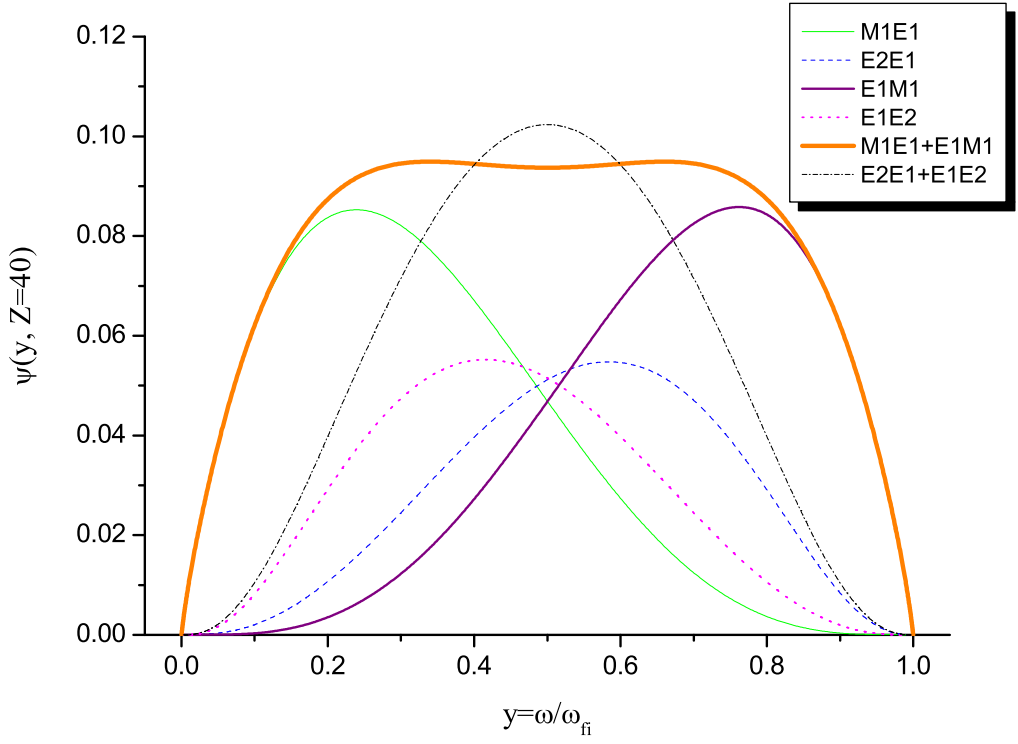}
\caption{Spectral distribution function $\psi (y,Z)$, defined by
  Eq. (\ref{esp}), of the $E1M1$ and   $E1E2$ contributions for the
  transition $2p_{1/2} \rightarrow 1s_{1/2}$ at $Z$=40. The variable
  $y=\omega/\omega_{fi}$ is the fraction of the photon energy
  carried by one of the two-photons.} 
\label{fig_2p}
\end{figure}

\begin{figure}[htb]
  \centering \includegraphics[clip=true,width=15cm]{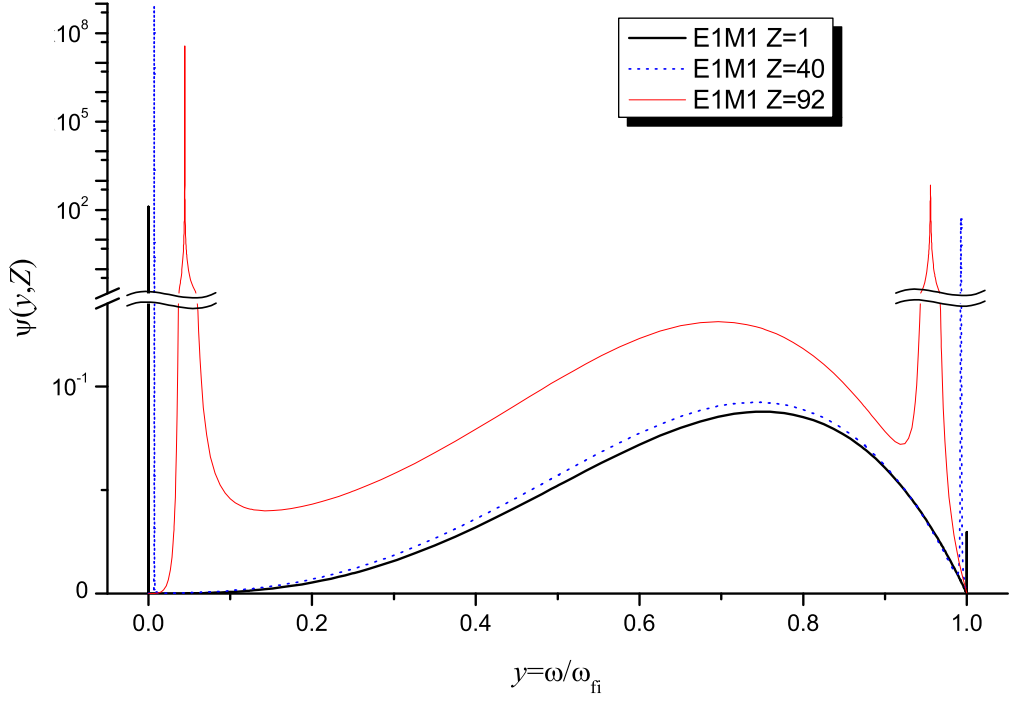}
\caption{Spectral distribution function $\psi (y,Z)$, defined by Eq.
  (\ref{esp}), of the $E1M1$ and contribution for the transition
  $2p_{3/2} \rightarrow 1s_{1/2}$ at $Z$=1, $Z$=40, $Z$=92. The
  variable $y=\omega/\omega_{fi}$ is the fraction of energy carried by one of the
  two-photons.}
\label{fig_2p3_2_E1M1}
\end{figure}

In Fig. \ref{fig_2p3_2_E1M1}, we plot the frequency distribution of
the multipole $E1M1$ contribution for the $2p_{3/2}\rightarrow 1s_{1/2}$
transition.  Along with the resonances, the shape of the curve is
similar to the one in Fig.~\ref{fig_2p} for the $2p_{1/2}\rightarrow
1s_{1/2}$ transition. In the $E1M1$ case, the resonance in the low-frequency
side occurs when the energy of one of the photons is equal to the
energy difference $E_{2p_{3/2}}-E_{2s_{1/2}}$, while the resonance in the
high-frequency side occurs when the energy of one of the photons is
equal to the energy difference $E_{2p_{3/2}}-E_{2p_{1/2}}$.

The list of the radiative corrections contributions for some states, which
were included in Eq.  (\ref{SjN_r}) to achieve an accuracy of at
least six digits are listed in Table \ref{Tab_radia_corre}.
The values for the real part of self-energy and vacuum polarization
were obtained from the MCDF code developed by Desclaux, Indelicato and
collaborators \cite{64,59,32}.  The level width, $\Gamma_{\nu}$, is
equal  to the sum of the one--photon partial level widths,
given by Eq. (\ref{one_pho}).

As seen from Eq. (\ref{rrii6}), by performing the integration of the
differential transition probabilities over energy of the emitted
photon we may finally obtain the total two-photon decay rates.
Eq.~(\ref{sum_h}) shows that these rates can be traced back to
$\mathfrak{h}$ functions.
In Table \ref{Tab_h_withcorr}, we list the sum of the terms
$\mathfrak{h}^{\textrm{LPA}}$ and $\mathfrak{h}^{\textrm{LPA}}_{1}$
given by Eq. (\ref{hh}) and Eq. (\ref{h111}), required for the
evaluation of the decay rates for transitions from bound states with
$n_{i}=3$ in the LPA,  including the most relevant multipoles,
radiative corrections and using $q=1$. The correspondent values
obtained in TLA are listed in Table \ref{Tab_h_without}.
By comparing the values in these two tables we conclude that they
differ less than 0.001\% for $Z=1$, 2.3\% for $Z=40$ and 10\% for
$Z=92$, which shows the importance of the radiative effects.

In Tables \ref{tab_2p3} and \ref{tab_3s}, we list the most relevant
multipole combinations included in the calculation of the two-photon
decay rate for the $2p_{3/2}\rightarrow 1s_{1/2}$ and
$3s_{1/2}\rightarrow 2s_{1/2}$ transitions.

\begin{figure}[htb]
\centering
\includegraphics[clip=true,width=15cm]{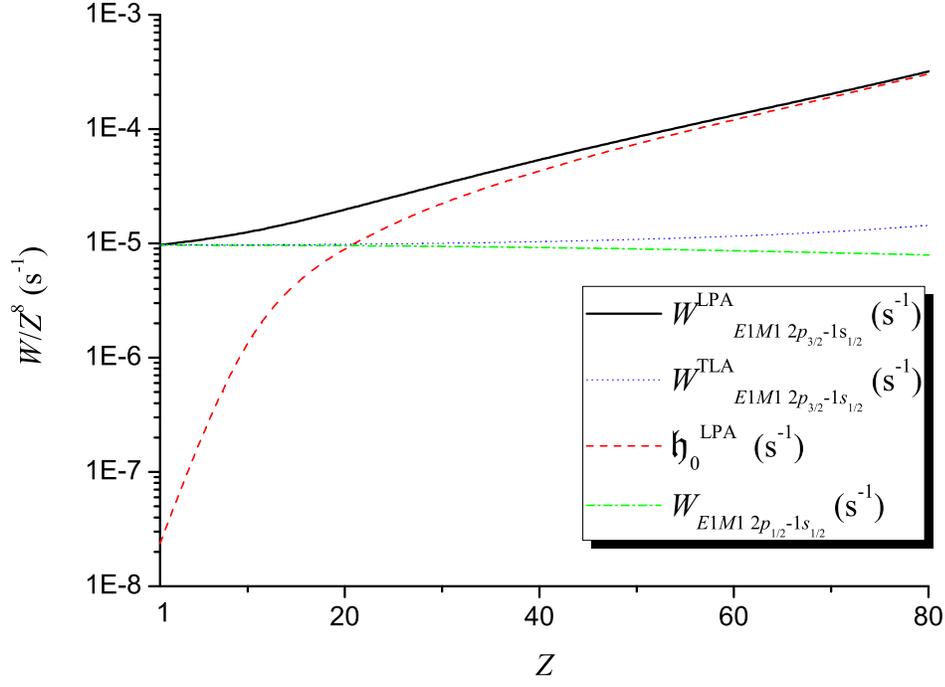}
\caption{Multipole combination $E1M1$ decay rate values $W_{E1M1}$, obtained in
  the LPA and TLA, as function of the atomic number $Z$ for the
  transitions $2p_{1/2}\rightarrow1s_{1/2}$ (dash-dot) and
  $2p_{3/2}\rightarrow1s_{1/2}$ (solid and dot lines). The cascade
  term in LPA is represented by the dash line.  }
\label{fig_2p3_2_Z}
\end{figure}

We notice that for $Z=1$ the decay rate values
of some multipole contributions, such as the $E1M1$ and the $E1E2$,
listed in Table \ref{tab_2p3}, are similar to the correspondent ones
for the transition $2p_{1/2} \rightarrow 1s_{1/2}$. Nevertheless, this
is not the case for $Z=40$ and $Z=92$.  This is due to the fact that
the energy separation between $2p_{3/2}$ and $2s_{1/2}$ increases with
$Z$ and, consequently, the decay rate contribution from the cascade
process also increases.
This aspect is also evident in Fig.~\ref{fig_2p3_2_Z}, where the
multipole combination $E1M1$ decay rate $W_{E1M1}$, obtained in the
LPA and TLA, is plotted as a function of the atomic number $Z$ for the
$2p_{1/2,3/2} \rightarrow 1s_{1/2}$ transitions.
  
The resonant behavior of the $2p_{3/2}\rightarrow 1s_{1/2}$ transition
is strongly suppressed for low $Z$ values. We notice that for lower
$Z$ values both solid (LPA) and dot (TLA) lines have similar values,
which is a consequence of the fact that nonresonant contribution
(related to integral of ``background"), in both transitions (M1E1 in
Figs.~\ref{fig_2p} and \ref{fig_2p3_2_E1M1}) is much higher than the
cascade term (dash line).
For higher values of $Z$, we notice that the solid line follows the
dash line.  This could be explained by the different $Z$ scaling of
the two contributions.  The ``background" scales as $Z^{8}$ and the
cascade term, given by $2p_{3/2}\rightarrow
2s_{1/2}\rightarrow1s_{1/2}$, scales as $Z^{10}$. The dash-dot (decay
rate of $2p_{1/2}-1s_{1/2}$) and dot lines are almost coincident in
low $Z$ region and diverge from about $Z=40$, which is an evidence of
the relativistic effects in the $np_{1/2}$ and $np_{3/2}$.

In Table \ref{Tab_23_n3} we report two-photon total decay rates for
transitions from initial level with $n_i=3$, obtained in the LPA
considering the most relevant multipole combinations in
Eq.~(\ref{tput}).  The results of Tung \textit{et al}, presented in
this table, were calculated using the analytical formulas described in
Ref. \cite{95}, which were obtained through the so called implicit
technique that describes the intermediate states by a differential
equation.

We restrict ourselves to list the two-photon decay rates obtained in
the LPA because in some cases they are very different from the TLA
ones, when the cascade term in Eq. (\ref{h_LPA_JA}) dominates.

One important aspect concerning total decay rates of resonant
transitions is the calculation of the nonresonant decay rate without
interference from resonant intermediate states.  Cresser \textit{et
  al} \cite{1993}, using a fourth-order perturbation term development,
obtained an expression similar to Eq.  (35.21) in Ref. \cite{12} where
the sum over the intermediate states considers only the states above
the initial one, avoiding in this way the resonant denominators, and
found the value $8.2197 \mbox{ s}^{-1}$ for the
$3s_{1/2}\rightarrow1s_{1/2}$ transition rate. Florescu \cite{2196},
using the same procedure, obtained the value $8.22581 \mbox{ s}^{-1}$.
The nonrelativistic limit of Eq. (\ref{eq_gera}) in the Coulomb gauge
gives the same expression as the one reported by Cresser \textit{et
  al} \cite{1993} and, consequently, the same result.

Jentschura \cite{2282} pointed out that Cresser \textit{et al}'s
procedure is not gauge invariant since in a second order evaluation
the sum over the complete spectrum of intermediate states is required
to have equivalence between two different gauges (more details are
given in appendix of Ref. \cite{3}).

Chluba and Sunyaev \cite{2723} developed another method to isolate the
nonresonant contribution. In their method, the sum over all the
intermediate states is split up in resonant and nonresonant states.
Although one can make conclusions for the difference between a pure
cascade process, i.e., considering only the resonant states with a
Lorentzian profile, and the two-photon emission given by all
intermediate states (resonant and nonresonant), the definition of a
nonresonant two-photon emission is unclear from a physical point of
view.

The values listed in Table \ref{Tab_h_without} were used to calculate
the nonresonant radiative corrections presented in Table
\ref{Tab_23_n3_nr} (setting $q=10^{-3}$). We notice that for the
$3s_{1/2}\rightarrow 1s_{1/2}$ transition the values calculated in
this work differ from the values obtained by Jentschura \cite{2285} by
0.01 \% for $Z=1$ and 0.1 \% for $Z=40$.

\begin{figure}[htb]
\centering
\includegraphics[clip=true,width=15cm]{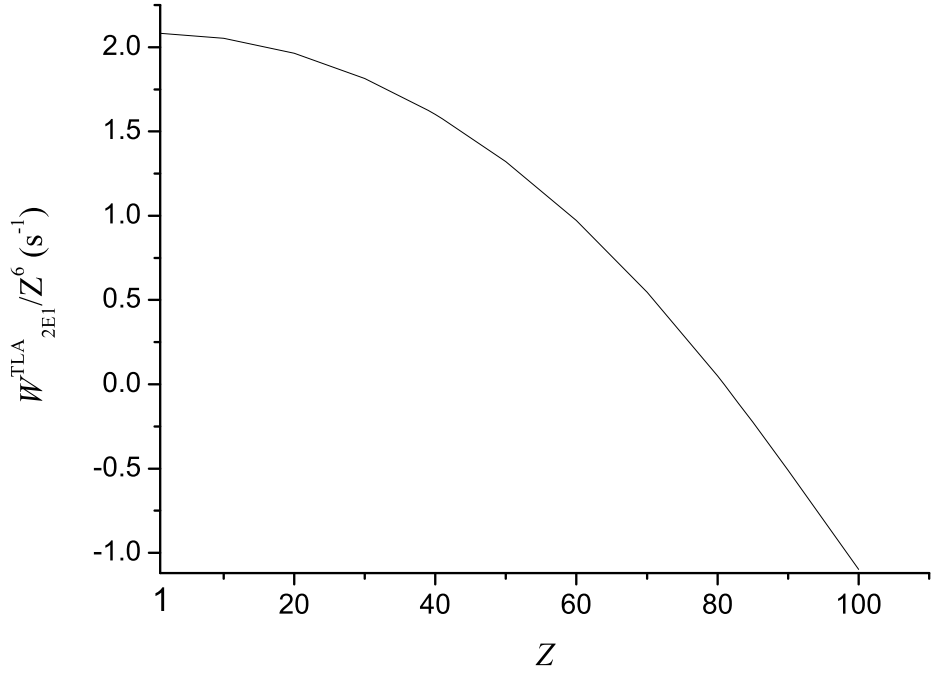}
\caption{Nonresonant correction of the multipole combination $2E1$ in
  the $3s_{1/2} \to 1s_{1/2}$ transition divided by $Z^{6}$ as
  function of the atomic number $Z$.}
\label{fig_NRC_Z}
\end{figure}

The reason for some values in Table \ref{Tab_23_n3_nr} being negative,
such as the $3s_{1/2}\rightarrow1s_{1/2}$ transition correction for
$Z=92$, is due to the evaluation of the two--loop self--energy, which
can be negative as any negative correction to the decay rates
\cite{2285}. In Fig. \ref{fig_NRC_Z}, we represent the values of the
nonresonant radiative correction for several values of atomic number.

%=====================================================================
%
\section{\label{Con}Conclusions}
%\label{Con}

By applying a finite basis set constructed from B-splines to solve the
Dirac equation, we have been able to calculate the decay rates in the
Line Profile and QED based on the two-loop self energy approaches for
all two-photon transitions from initial states with $n=2$ and 3 for a
set of hydrogenlike ions with nuclear charge ranging from $Z=1$ to
$Z=92$. In these calculations the most significant multipoles
contributions were considered, such as the $2E1$, $E1M1$, $2M1$, etc.
We have also studied the spectral distributions of several
transitions, which exhibit specific structures, such as resonances and
transparencies. The latter reveal that two-photon emission is not
possible at certain frequencies.  The numerical results obtained in
this work are in good agreement with other nonrelativistic and
relativistic theoretical results.

The QED approach gives a better contribution for a pure coherent
nonresonant two-photon emission than Cresser's and Chluba's methods,
not only because it is derived from physical arguments, but also due
to the fact that it can be obtained from the Line Profile approach by
removing the cascade process and setting the radiative corrections to
zero. Therefore, it is a useful technique in theoretical evaluations
that require a coherent two-photon decay rate rather than the sum of
this term along with the sequential one photon decay rate (cascade
process).

We conclude that the Line Profile approach is the most suitable for
comparison with experimental results since it includes the terms
associated with cascade process as well as radiative corrections. 

We end this conclusion by emphasizing that the method of integration
used to obtain one electron decay rates (in both approaches and for
both nonresonant and resonant transitions) can be adapted perfectly to
ions with two or three electrons.

%
%======= ACKNOWLEDGMENTS =============================================
%
\begin{acknowledgments}
%\section*{Acknowledgments}
%
%  
  This research was supported in part by FCT project
  POCTI/FAT/44279/2002 and POCTI/0303/2003(Portugal), financed by the
  European Community Fund FEDER, by the French-Portuguese
  collaboration (PESSOA Program, Contract n$^\mathrm{o}$ 441.00), and
  by the Ac{\c c}{\~o}es Integradas Luso-Francesas (Contract
  n$^\mathrm{o}$ F-11/09).  The work of A.S. was supported by the
  Helmholtz Gemeinschaft (Nachwuchsgruppe VH--NG--421). Laboratoire
  Kastler Brossel is ``Unit\'e Mixte de Recherche du CNRS, de l' ENS
  et de l'UPMC n$^{\circ}$ 8552''. P. Indelicato acknowledges the
  support of the Helmholtz Allianz Program of the Helmholtz
  Association, contract HA-216 "Extremes of Density and Temperature:
  Cosmic Matter in the Laboratory". P. Amaro acknowledges the support
  of the FCT, contract SFRH/BD/37404/2007.

\end{acknowledgments}

%
%======= APPENDIX ====================================================
%
\appendix

\section{Appendix}

In order to make the task of deriving the matrix elements less
cumbersome, further simplifications can be done in the matrix elements
(Eq.~(\ref{M_al_bet1}) and (\ref{M_min})) by noticing that the
longitudinal part of the operator $\widetilde{a}^{(\lambda)}_{LM}$
\cite{2280},
\begin{equation}
\left(\widetilde{a}^{(-1)}_{LM}\right)_
{||}=\frac{c}{i\omega}{\bf \alpha}\cdot{\bf \nabla}\phi_{L,M},
\end{equation}
can be writen using a commutation relation as
\begin{eqnarray}
\left(\widetilde{a}^{(-1)}_{LM}\right)_
{||}&=&\frac{c}{i\omega}\left[H_{D},\phi_{L,M}\right],
\label{comut}
\end{eqnarray}
where $H_{D}$ stands for the Dirac Hamiltonian and $\phi_{L,M}$ are
the components of the spherical tensor of rank $L$ resulting from the
multipole expansion of the potential $A^{*}_{j}$.  The reduction of
Eq.~(\ref{comut}) to radial integrals along with the scalar term of the
potential $A^{*}_{j}$, lead to the following expression for the radial
element matrix $\overline{M}_{f,i}^{(-1,L)}$
\begin{equation}
\overline{M}_{f,i}^{(-1,L)}=G(2L+1)\left(\frac{\omega+\omega_{fi}}{\omega}\right)J^{\left(
    L\right) }, 
\label{M_min_sim} 
\end{equation}
where $\omega_{fi}$ is the energy of the one-photon transition.  This
term is gauge independent for one-photon, as demonstrated by Grant
\cite{6}, since $\omega_{fi}=-\omega$. Considering Eq. (\ref{comut}),
 the radial matrix element $\overline{M}_{f,i}^{(1,L)}$ can also be rewritten as
\begin{eqnarray}
\overline{M}_{f,i}^{(1,L)}&=&\frac{(2L+1)}{\sqrt{L(L+1)}}
\left[-(\kappa_{f}-\kappa_{i})I^{+}_{L-1} \right. \nonumber  \\ 
& + & LI^{-}_{L-1} + \left.L\frac{\omega_{fi}}{\omega}J^{L}\right]
\label{M_1_sim}.
\end{eqnarray}

The explicit expressions of the derivatives of the matrix elements
(Eq.~(\ref{M_1_sim}), Eq.~(\ref{M_al_bet}) and Eq.~(\ref{M_min_sim})) are
given by
\begin{eqnarray}
\frac{d}{d\omega }\left[ \overline{M}_{f,i}^{(1,L)}(\omega )\right] &=&\frac{
2L+1}{\sqrt{L(L+1)}}\left[ (\kappa _{f}-\kappa _{i})\left(
  I_{L}^{\textrm{R}}+\frac{I_{L-1}^{+}}{\omega }\right) \right. \nonumber\\ 
&-&\left.L\frac{I_{L-1}^{-}}{\omega }-\frac{\omega _{fi}%
}{\omega ^{2}}L\left( L+2\right) J^{\left( L\right) }\right] ,
\end{eqnarray}
\begin{equation}
\frac{d}{d\omega }\left[ \overline{M}_{f,i}^{(0,L)}(\omega )\right] =\frac{
2L+1}{\sqrt{L(L+1)}}\left( \kappa _{f}+\kappa _{i}\right) \left[ I_{L-1}^{\textrm{R}}-
\frac{\left( L+1\right) }{\omega }I_{L}^{+}\right] ,
\end{equation}
and
\begin{eqnarray}
\frac{d}{d\omega }
\left[ \overline{M}_{f,i}^{(-1,L)}(\omega )\right] &=&G(2L+1)
\left\{ 
    \left( \frac{\omega +\omega_{fi}}{\omega }\right)  J_{\textrm{R}}^{\left(
        L-1\right) } 
\right. \nonumber\\
&-& 
\left. 
  \frac{1}{\omega ^{2}}\left[ \left( L+1\right) \omega
  +\left(L+2\right) \omega _{fi} \right] J^{\left( L\right) }
\right\} , 
\nonumber\\
\end{eqnarray}
where the integrals $J_{\textrm{R}}^{\left( L\right) }$ and $I_{L}^{\textrm{R}}$ are
defined by
\begin{eqnarray}
I_{L}^{\textrm{R}} &=&\frac{1}{c}\int_{0}^{\infty }\left(
P_{f}Q_{i}+P_{i}Q_{f}\right) rj_{L}\left( \frac{\omega r}{c}\right) dr, \\
J_{\textrm{R}}^{\left( L\right) } &=&\frac{1}{c}\int_{0}^{\infty }\left(
P_{f}P_{i}+Q_{f}Q_{i}\right) rj_{L}\left( \frac{\omega r}{c}\right) dr.
\end{eqnarray}

Using these expressions, along with the definitions given by
Eqs.~(\ref{aaa01}), we were able to obtain the coefficients
$a_{0}^{j}$ and $a_{1}^{j}$ listed in Table \ref{tab_a0_a1} for the
transition $3s_{1/2} \rightarrow 1s_{1/2}$ for ions with $Z=1$, 40 and
92.
%
%
%========= BIBLIOGRAPHY ==============================================
%

%
% BibTeX users please use
% \bibliographystyle{}

%\bibliography{jps}
%
\newpage
%

%========= TABLES ==================================================== 
%

%TABLE 1
\begin{table}
\caption{\label{tab_2p}Multipole contributions included in the present
  calculation of the total two-photon rate for the
  $2p_{1/2}\rightarrow 1s_{1/2}$   transition. Comparison between the
  values obtained in this work (Eq.~\ref{rrii6}) and  other theoretical values. Powers
  of ten are given in parentheses. 
}
\begin{ruledtabular}
\begin{tabular}{llll}
%\hline\hline
Multipoles & \multicolumn{3}{c}{Contribution (s$^{-1}$)} \\ 
       & $Z=1$ & $Z=40$  & $Z=92$ \\ \hline
$E1M1$ & $9.676654(-6)$  & $6.027323(7)$  & $3.863302(10) $ \\ 
       & $9.667(-6)$\footnotemark[1]     & $6.020(7)$\footnotemark[1]   & $3.859(10)$\footnotemark[1]  \\ 
       & $9.677(-6)$\footnotemark[2]     & $6.341(7)$\footnotemark[2]    & $4.966(10)$\footnotemark[2] \\ 
\cline{2-4}
$E1E2$ & $6.61179(-6)$   & $4.092020(7)$ & $2.358404(10) $ \\ 
       & $6.605(-6)$\footnotemark[1]     & $4.088(7)$\footnotemark[1]    & $2.357(10)$ \footnotemark[1] \\ 
       & $6.673(-6)$\footnotemark[2]     & $4.374(7)$\footnotemark[2]    & $3.425(10)$\footnotemark[2] \\ 
\cline{2-4}
$M1M2$ & $3.827877(-17)$  & $5.602320(2)$ & $7.689142( 6) $ \\ 
\cline{2-4}
$E2M2$ & $9.385470(-17)$ & $1.521687(3)$ & $2.834065( 7) $ \\ 
\cline{2-4}
$E2E3$ & $4.095985(-18)$ & $6.608612(1)$ & $1.177403( 6) $ \\ 
\\
Total  & $1.628845(-5)$  & $1.01195(8)$ & $6.225309(10)$
\\ 
%\hline
%\hline
\end{tabular}
\end{ruledtabular}
\footnotetext[1]{Labzowsky {\it et al} \cite{1418}} 
\footnotetext[2]{Labzowsky {\it et al} \cite{1907}} 
\end{table}
%
%
%========= TABLES ==================================================== 
% 

%TABLE 2
\begin{table}
\caption{\label{Tab_23_n2}Total two-photon decay rates (s$^{-1}$) for the transitions
  $2s_{1/2}\rightarrow1s_{1/2}$ and $2p_{1/2}\rightarrow1s_{1/2}$. Comparison between
  the values obtained in this work (Eq.~\ref{tput})  and other theoretical
  values. Powers of ten are given in parentheses. 
}
\begin{ruledtabular}
\begin{tabular}{llll}
%\hline\hline
& \multicolumn{3}{c}{Total decay rate (s$^{-1}$)} \\ 
\hline
$Z=1$ & $f\backslash i$ & $2s_{1/2}$ & $2p_{1/2}$   \\ 
\cline{2-4}
& $1s_{1/2}$ & 8.229059               & 1.628845(-5)              \\
&            & 8.2202 \footnotemark[1] & 1.6272(-5) \footnotemark[1] \\
&            &                       & 1.6350(-5)\footnotemark[2]  \\
\hline
$Z=40$ & $f\backslash i$ & $2s_{1/2}$ & $2p_{1/2}$  \\ 
\cline{2-4}
& $1s_{1/2}$ & 3.198851(10)                & 1.01195(8)             \\
&            & 3.1954(10) \footnotemark[1]  & 1.010(8)\footnotemark[1]  \\
&            &                           & 1.071(8)\footnotemark[2]   \\
\hline
$Z=92$ & $f\backslash i$ & $2s_{1/2}$ & $2p_{1/2}$  \\ 
\cline{2-4}
& $1s_{1/2}$ & 3.835978(12)               & 6.225309(10)            \\
&            & 3.8216(12) \footnotemark[1] & 6.216(10)\footnotemark[1] \\ 
&            &                          & 8.391(10)\footnotemark[2] \\ 
%\hline\hline
\end{tabular}
\end{ruledtabular}

\footnotetext[1]{Labzowsky {\it et al} \cite{1418}}
\footnotetext[2]{Labzowsky {\it et al} \cite{1907}} 
\end{table}
%           
%
%========= TABLES ==================================================== 
% 

%TABLE 3
\begin{table}[tbp]
\caption{\label{tab_transp}Transparencies for several two-photon   transitions. The
  variable $y=\omega_{1}/\omega_{fi}$ is the fraction of the photon
    energy carried by one of the two-photons. 
}
\begin{ruledtabular}
\begin{tabular}{llll}
%\hline\hline
Transition & $y$ $(Z=1)$ & $y$ $(Z=40)$ & $y$ $(Z=92)$ \\ 
\hline 
$3s_{1/2}\rightarrow 1s_{1/2}$ & 0.780267 & 0.77628 & 0.7518 \\
                   & 0.7803 \footnotemark[1] \\
                   & 0.7802 \footnotemark[2] \\
                   & 0.7803 \footnotemark[3] \\
$4s_{1/2}\rightarrow 1s_{1/2}$ & 0.737322 & 0.73273 & 0.7034 \\ 
                   & 0.7373 \footnotemark[1] \\
                   & 0.7373 \footnotemark[3] \\
$6s_{1/2}\rightarrow 1s_{1/2}$ & 0.7032201 & 0.70497 & 0.6725 \\
                   & 0.7098 \footnotemark[1] \\
                   & 0.7079 \footnotemark[2] \\
                   & 0.7098 \footnotemark[3] \\
%\hline
%\hline
\end{tabular}
\end{ruledtabular}

\footnotetext[1]{Florescu {\it et al} \cite{2284}}
\footnotetext[2]{Quattropani {\it et al} \cite{144}}
\footnotetext[3]{Tung {\it et al} \cite{95}}
\end{table}
%
%
%
%========= TABLES ==================================================== 
% 

%TABLE 4
\begin{table*}
\caption{\label{Tab_radia_corre}Radiative corrections for several states in a.u. Powers of ten are given in parentheses.
} 
\begin{ruledtabular}
\begin{tabular}{llllllll}
%\hline\hline
& \multicolumn{3}{c}{Radiative terms} (a. u.) \\
\hline
$Z=1$ & state & $2s_{1/2}$ & $2p_{1/2}$ & $2p_{3/2}$ & $3s_{1/2}$& $3p_{1/2}$& $3p_{3/2}$ \\  
\cline{2-8}
& Re [SE]+VP$=\Delta E_{n}$ & 1.5867(-7)               & -1.9542(-9)  & 1.9095(-9)&  4.7376(-08)&-5.5003(-10)&6.4103(-10)           \\
& Im [SE]$=\Gamma_{n}/2$   & 1.9905(-16)              & 1.5162(-8)   & 1.5162(-8)&  1.5281(-10)&4.5911(-9)&4.5911(-9)        \\
\hline
$Z=40$ & state & $2s_{1/2}$ & $2p_{1/2}$ & $2p_{3/2}$ & $3s_{1/2}$& $3p_{1/2}$& $3p_{3/2}$  \\  
\cline{2-8}
& Re [SE]+VP$=\Delta E_{n}$ & 8.6991(-2)               & -1.4913(-3) & 7.0905(-3)& 2.6401(-2)&-1.3587-4)& 2.3058(-3)    \\
& Im [SE]$=\Gamma_{n}/2$   & 1.4690(-6)               & 3.9208(-2)  &3.8037(-2) & 4.6373(-4)& 1.1689(-2)& 1.1627(-2)      \\
\hline
$Z=92$ & state & $2s_{1/2}$ & $2p_{1/2}$ & $2p_{3/2}$ & $3s_{1/2}$& $3p_{1/2}$& $3p_{3/2}$  \\  
\cline{2-8}
& Re [SE]+VP$=\Delta E_{n}$ & 1.7995                   & 2.4965(-1)& 3.2252(-1)&5.7911(-01)& 9.3654(-2)& 1.102(-1)     \\
& Im [SE]$=\Gamma_{n}/2$   & 4.7468(-3)               & 1.1417    & 9.5531(-1)& 2.7061(-2)& 3.1125(-1)& 3.0444(-1)       \\
%\hline\hline
\end{tabular}
\end{ruledtabular}
\end{table*}
%
%========= TABLES ==================================================== 
% 
%

%TABLE 5
\begin{table*}
\caption{\label{Tab_h_withcorr}Sum of the terms $\mathfrak{h}^{\textrm{LPA}}$ and
  $\mathfrak{h}^{\textrm{LPA}}_{1}$, given by Eqs.~(\ref{hh}) and
  \ref{h111}), respectively, for transitions
  from bound states with $n_{i}=3$.  This values were obtain using the
    radiative corrections of Table \ref{Tab_radia_corre} and
    $q=1$. Powers of ten are given in parentheses. 
%\tiny
} 
\begin{ruledtabular}
\begin{tabular}{lllllll}
%\hline\hline
%& \multicolumn{6}{c}{Total decay rate, $\omega $ (s$^{-1}$)} \\ 
%\hline
$Z=1$ & $f\backslash i$ & $3s_{1/2}$ & $3p_{1/2}$ & $3p_{3/2}$ & $3d_{3/2}$ & $3d_{5/2}$ \\ 
\cline{2-7}
& $1s_{1/2}$ & 2.342758   & 3.966941(-6)  & 3.230044(-6)  & 3.706845 & 3.706854  \\
& $2s_{1/2}$ & 6.452435(-2) & 4.925801(-8) & 4.926319(-8) & 7.762447(-4)  & 7.750004(-4) \\
& $2p_{1/2}$ & 2.894796(-8) & 4.660148(-2) & 4.414498(-4) & 3.890718(-8)  & 3.049476(-9) \\ 
& $2p_{3/2}$ & 5.789457(-8) & 8.832671(-4)  & 4.704893(-2) & 1.326766(-8)  & 4.912187(-8) \\
% --------
\hline
$Z=40$ & $f\backslash i$ & $3s_{1/2}$ & $3p_{1/2}$ & $3p_{3/2}$ & $3d_{3/2}$ & $3d_{5/2}$ \\ 
\cline{2-7}
& $1s_{1/2}$ & 7.813052(9)   & 2.872804(7)  & 2.002016(7)  & 1.436349(10) & 1.441446(10) \\ 
& $2s_{1/2}$ & 2.247363(8)   & 2.812583(5)  & 3.395829(5)  & 6.369235(6)  & -1.183514(6)  \\ 
& $2p_{1/2}$ & 1.756667(5)   & 1.797278(8)  & 3.117290(6)  & 3.012475(5)  & 2.478360(4)  \\ 
& $2p_{3/2}$ & 3.370367(5)   & 8.545466(6)  & 2.168435(8)  & 8.878266(4)  & 3.231816(5)  \\ 
\hline
$Z=92$ & $f\backslash i$ & $3s_{1/2}$ & $3p_{1/2}$ & $3p_{3/2}$ & $3d_{3/2}$ & $3d_{5/2}$ \\ 
\cline{2-7}
& $1s_{1/2}$ & -5.316586(10) & 3.647738(10) & 2.048248(11)  & 1.646524(12) & 1.637676(12)  \\ 
& $2s_{1/2}$ & 7.443297(9) & 1.683735(8)  & 6.928218(8)   & 1.135193(10)   & 6.013487(9)  \\ 
& $2p_{1/2}$ & 9.548939(7)  & 2.314262(10) & 3.637390(9)   & 4.841719(8)   & 1.534768(7)  \\ 
& $2p_{3/2}$ & 1.554405(8)  & 5.312377(9)  & 5.376493(10)  & 1.660640(8)   & 3.122434(8)   \\ 
\\
%\hline\hline
\end{tabular}
\end{ruledtabular}
\end{table*}
%\cleardoublepage
%
%========= TABLES ==================================================== 
% 
%

%TABLE 6
\begin{table*}
\caption{\label{Tab_h_without}Sum of the terms $\mathfrak{h}^{\textrm{TLA}}$ and
  $\mathfrak{h}^{\textrm{TLA}}_{1}$, given by Eqs. (\ref{hhja}) and
  (\ref{h1_ja}), respectively, for transitions from
  bound states with $n_{i}=3$.  This values were obtain without
    radiative corrections, using  $q=10^{-2}$ and following  and using
    Jentschura's approach. Powers of ten are given in parentheses.
%\tiny
} 
\begin{ruledtabular}
\begin{tabular}{lllllll}
%\hline\hline
%& \multicolumn{6}{c}{Total decay rate, $\omega $ (s$^{-1}$)} \\ 
%\hline
$Z=1$ & $f\backslash i$ & $3s_{1/2}$ & $3p_{1/2}$ & $3p_{3/2}$ & $3d_{3/2}$ & $3d_{5/2}$ \\ 
\cline{2-7}
& $1s_{1/2}$ & 2.342751     & 3.966941(-6)  & 3.230044(-6)  & 3.706845     & 3.706854  \\
& $2s_{1/2}$ & 6.452428(-2) & 4.925765(-8) & 4.926337(-8) & 7.762447(-4)  & 7.750004(-4) \\
& $2p_{1/2}$ & 2.894793(-8) & 4.660148(-2) & 4.414498(-4) & 3.890718(-8)  & 3.049476(-9) \\ 
& $2p_{3/2}$ & 5.789453(-8) & 8.832670(-4)  & 4.704892(-2) & 1.326766(-8)  & 4.912188(-8) \\
% --------
\hline
$Z=40$ & $f\backslash i$ & $3s_{1/2}$ & $3p_{1/2}$ & $3p_{3/2}$ & $3d_{3/2}$ & $3d_{5/2}$ \\ 
\cline{2-7}
& $1s_{1/2}$ & 7.799875(9)  & 2.872822(7) &  2.000453(7)   & 1.436350(10) & 1.441394(10) \\ 
& $2s_{1/2}$ & 2.246945(8)   & 2.802618(5)  & 3.400453(5)  & 6.369529(6)  & -1.180529(6)  \\ 
& $2p_{1/2}$ & 1.756241(5)   & 1.797280(8)  & 3.116223(6)  & 3.080467(5)  & 2.477876(4)  \\ 
& $2p_{3/2}$ & 3.369973(5)   & 8.514049(6)  & 2.167575(8)  & 8.872957(4)  & 3.233897(5)  \\ 
\hline
$Z=92$ & $f\backslash i$ & $3s_{1/2}$ & $3p_{1/2}$ & $3p_{3/2}$ & $3d_{3/2}$ & $3d_{5/2}$ \\ 
\cline{2-7}
& $1s_{1/2}$ & -5.843073(10) & 3.632557(10) & 2.037502 (11)  & 1.648325(12) & 1.637187(12)  \\ 
& $2s_{1/2}$ & 7.703355(9) & 1.624996(8)  & 7.340967(8)   & 1.135843(10)   & 6.400342(9)  \\ 
& $2p_{1/2}$ & 9.552389(7)  & 2.313307(10) & 3.630875(9)   & 4.857978(8)   & 1.5282890(7)  \\ 
& $2p_{3/2}$ & 1.556356(8)  & 5.270175(9)  & 5.351656(10)  & 1.655470(8)   & 3.134378(8)   \\ 
\\
%\hline\hline
\end{tabular}
\end{ruledtabular}
\end{table*}  
%\newpage               
%
%========= TABLES ==================================================== 
% 
%

%TABLE 7
\begin{table}[tbp]
\caption{\label{tab_2p3}Same as Table \protect\ref{tab_2p} for the transition
  $2p_{3/2}\rightarrow 1s_{1/2}$. Powers of ten are given in
  parentheses. 
}
\begin{ruledtabular}
\begin{tabular}{clll}
%\hline\hline
Multipoles & \multicolumn{3}{c}{Contribution (s$^{-1}$)} \\ 
\multicolumn{1}{l}{} & $Z=1$ & $Z=40$ & $Z=92$ \\ \hline
\multicolumn{1}{l}{$E1M1$} & $9.700994(-6)$ & $3.547078\left( 8\right) $ & $2.93820\left( 12\right) $ \\ 
\multicolumn{1}{l}{$E1E2$} & $6.612242(-6)$ & $4.597372(7)$ & $1.113120\left( 11\right) $ \\ 
\multicolumn{1}{l}{$E1M3$} & $1.761532(-16)$ & $3.264236(3)$ & $1.216566\left( 8\right) $ \\ 
\multicolumn{1}{l}{$M1M2$} & $2.450145(-15)$ & $4.265433(4)$ & $1.129082\left( 9\right) $ \\ 
\multicolumn{1}{l}{$E2M2$} & $7.227055(-17)$ & $1.337970(3)$ & $4.923824\left( 7\right) $ \\ 
\multicolumn{1}{l}{$E2E3$} & $4.096369(-18)$ & $7.718214(1)$ & $1.216566\left( 8\right) $ \\ 
\\
\multicolumn{1}{l}{Total} & $1.631323(-5)$ & $4.007255\left( 8\right) $ & $3.050699(12)$  \\ 
%\hline
%\hline
\end{tabular}%
\end{ruledtabular}
\end{table}
 
%
%========= TABLES ==================================================== 
% 

%TABLE 8
\begin{table}[tbp]
\caption{\label{tab_3s}Same as Table \protect\ref{tab_2p} for the transition $
3s_{1/2}\rightarrow 2s_{1/2}$. Powers of ten are given in
    parentheses. 
}
\begin{ruledtabular}
\begin{tabular}{llll}
%\hline\hline
Multipoles & \multicolumn{3}{c}{Contribution (s$^{-1}$)} \\ 
& $Z=1$ & $Z=40$ & $Z=92$ \\ \hline
$2E1$  & $6.452436(-2)$  & $3.167606(8)$  & $1.850546\left( 12\right) $ \\ 
$E1M2$ & $6.725935(-14)$ & $1.150220(3)$  & $3.359725\left( 8\right) $ \\ 
$2M1$  & $1.038556(-14)$ & $1.241685(2)$  & $1.031596\left( 6\right) $ \\ 
$2E2$  & $1.456030(-14)$ & $1.584123(2)$  & $8.482772\left( 5\right) $ \\ 
$2M2$  & $1.901242(-27)$ & $6.068107(-5)$ & $4.479357\left( 3\right) $ \\ 
\\
Total  & $6.452436(-2)$  & $3.167620(8)$  & $1.850884\left( 12\right) $ \\ 
%\hline
%\hline
\end{tabular}%
\end{ruledtabular}
\end{table}
%\pagebreak
%\cleardoublepage
%
%
%========= TABLES ==================================================== 
% 

%TABLE 9
\begin{table*}
\caption{\label{Tab_23_n3}Total two-photon decay rates (s$^{-1}$)  in
  the LPA, given by  Eq. (Eq.~\ref{tput}), for transitions from bound states with
  $n_{i}=3$.  Powers of ten are given in parentheses.  
%\tiny
} 
\begin{ruledtabular}
\begin{tabular}{lllllll}
%\hline\hline
& \multicolumn{6}{c}{Total decay rate (s$^{-1}$)} \\ 
\hline
$Z=1$ & $f\backslash i$ & $3s_{1/2}$ & $3p_{1/2}$ & $3p_{3/2}$ & $3d_{3/2}$ & $3d_{5/2}$ \\ 
\cline{2-7}
% -----
& $1s_{1/2}$ & 6.382020(6)  & 3.431055(1)   & 3.431043(1)  & 7.213121(7)   & 7.212970(7) \\
& $2s_{1/2}$ & 6.452436(-2) & 4.925806(-8)  & 4.965339(-8) & 7.762774(-4)  & 7.751032(-4) \\
&            & 6.4527(-2)\footnotemark[1] &    &              & 7.7589(-4)\footnotemark[1] &  \\
& $2p_{1/2}$ & 2.894796(-8) & 4.6601485(-2) & 4.414514(-4) & 3.890719(-8)  & 3.070354(-9)  \\ 
&            &              & 4.7484(-2)\footnotemark[1] \\
& $2p_{3/2}$ & 5.789457(-8) & 8.832671(-4)  & 4.704893(-2) &  1.326768(-8) & 4.912603(-8) \\
\hline
$Z=40$ & $f\backslash i$ & $3s_{1/2}$ & $3p_{1/2}$ & $3p_{3/2}$ & $3d_{3/2}$ & $3d_{5/2}$ \\ 
\cline{2-7}
& $1s_{1/2}$ & 1.940069(13)  & 1.465902(11) & 1.459393(13) & 1.909365(14) & 1.846466(14) \\ 
& $2s_{1/2}$ & 3.167620(8)   & 1.249212(6)  & 6.204266(6)  & 4.239616(8)  & 1.256275(9)  \\ 
& $2p_{1/2}$ & 2.488771(5)   & 1.797286(8)  & 2.342989(7)  & 4.767668(5)  & 1.268419(6)  \\ 
& $2p_{3/2}$ & 3.370367(5)   & 8.545466(6)  & 2.648671(8)  & 4.278490(5)  & 3.734221(5)  \\ 
\hline
$Z=92$ & $f\backslash i$ & $3s_{1/2}$ & $3p_{1/2}$ & $3p_{3/2}$ & $3d_{3/2}$ & $3d_{5/2}$ \\ 
\cline{2-7}
& $1s_{1/2}$ & 1.067206(15) & 5.649957(13) & 4.187334(13)  & 6.170663(15)  & 5.175621(15)  \\ 
& $2s_{1/2}$ & 1.850884(12) & 4.581582(10) & 1.219838(11)  & 5.062691(12)  & 1.123469(13)  \\ 
& $2p_{1/2}$ & 8.893987(9)  & 2.336202(10) & 2.576780(11)  & 9.697069(9)   & 5.543826(10)  \\ 
& $2p_{3/2}$ & 1.554405(8)  & 5.312377(9)  & 1.059702(12)  & 1.959473(10)  & 1.239531(9)   \\ 
\\
%\hline\hline
\end{tabular}
\end{ruledtabular}

\footnotetext[1]{Tung {\it et al} \cite{95}}
\end{table*}
%
% 
%========= TABLES ==================================================== 
% 

%\pagebreak
%\ 
%\newpage
%\pagebreak
%\newpage
%\ 

%TABLE 10
\begin{table*}[tbp]
\caption{\label{Tab_23_n3_nr}Total nonresonant two-photon correction (s$^{-1}$)
  in the TLA , given by Eq.~(\ref{tput}), for  transitions from bound
    states with $n_{i}=3$. Comparison between the values obtained
    in this work and other theoretical values. Powers of ten are
    given in parentheses.  
%\tiny
} 
\begin{ruledtabular}
\begin{tabular}{lllllll}
%\hline\hline
& \multicolumn{6}{c}{Total nonresonant correction (s$^{-1}$)} \\ 
\hline
$Z=1$ & $f\backslash i$ & $3s_{1/2}$ & $3p_{1/2}$ & $3p_{3/2}$ & $3d_{3/2}$ & $3d_{5/2}$ \\ 
\cline{2-7}
% -----
& $1s_{1/2}$ & 2.082562   & 2.981766(-6) & 2.98676(-6) & 1.042768     & 1.042835   \\
&            & 2.082853 \footnotemark[1] &    &              &
1.042896 \footnotemark[2]        &  \\
% ------
& $2s_{1/2}$ & 6.452428(-2) & 4.925721(-8) & 4.926293(-8) & 7.762407(-4) & 7.749962(-4) \\
&            & 6.4530(-2) \footnotemark[1] &    &              &                          &  \\
% -----
& $2p_{1/2}$ & 2.894793(-8) & 4.6601486(-2) & 4.414498(-4) & 3.890718(-8) & 3.049476(-9)  \\ 
% ------
& $2p_{3/2}$ & 5.789453(-8) & 8.832670(-4) & 4.704892(-2) &  1.326767(-8) & 4.912188(-8) \\
%&            &              &               & 4.7484(-2)$^{\text b}$ \\
% --------
\hline
$Z=40$ & $f\backslash i$ & $3s_{1/2}$ & $3p_{1/2}$ & $3p_{3/2}$ & $3d_{3/2}$ & $3d_{5/2}$ \\ 
\cline{2-7}
& $1s_{1/2}$ & 6.560351(9)   & 2.224659(7) & 1.885681(7) & 3.456276(9) & 3.874677(9)   \\ 
& $1s_{1/2}$ & 6.554(9) \footnotemark[3]     &                &               &             &               \\ 
& $2s_{1/2}$ & 2.245669(8)   & 2.793230(5)  & 3.395425(5)  & 5.920457(6)  & -2.764917(6)  \\ 
& $2p_{1/2}$ & 1.755215(5)   & 1.797280(8)  & 3.088316(6)  & 3.078375(5)  & 2.349043(4)  \\ 
& $2p_{3/2}$ & 3.369973(5)   & 8.514049(6)  & 2.166915(8)  & 8.842254(4)  & 3.233689(5)  \\ 
\hline
$Z=92$ & $f\backslash i$ & $3s_{1/2}$ & $3p_{1/2}$ & $3p_{3/2}$ & $3d_{3/2}$ & $3d_{5/2}$ \\ 
\cline{2-7}
& $1s_{1/2}$ & -3.842113(11) & 2.891626(10)  & 7.976296(10)   & 8.916260(10)  & 3.271857(11)  \\ 
& $2s_{1/2}$ & 5.570205(9)   & 1.258571(8) & 7.132981(8)  & 7.613467(9)  & -5.085413(9)  \\ 
& $2p_{1/2}$ & 8.535648(7)  & 2.313315(10) & 3.386706(9)  & 4.767493(8)   & -5.866950(7)   \\ 
& $2p_{3/2}$ & 1.556356(8)  & 5.270175(9)  & 5.253150(10)  & 1.539227(8)  & 3.127867(8)   \\ 
\\
%\hline\hline
\end{tabular}
\end{ruledtabular}

\footnotetext[1]{Jentschura \cite{2285}}
\footnotetext[2]{Jentschura \cite{2635}}
\footnotetext[3]{Jentschura \cite{2491}}
\end{table*}
\ 
%
% 
%
%========= TABLES ==================================================== 
%
%TABLE 11
%\begin{table}[tbp]
\begin{table}
\caption{\label{tab_a0_a1}Values of the  coefficients $a_{0}^{j}$
  ($s^{-2}$) and $a_{1}^{j}$ ($s^{-1}$), given by Eqs.~(\ref{aaa01}),
    of the transition  $3s_{1/2} \rightarrow 1s_{1/2}$ for several
    values of $Z$. Powers of ten are given in parentheses. 
\protect}
\begin{ruledtabular}
\begin{tabular}{llll}
%\hline\hline
       & $Z=1$ & $Z=40$  & $Z=92$ \\ \hline
$a_{0}^{1/2}$ & $5.081524(-3)$    & $3.554138(10)$  & $3.793201(13)$ \\ 
$a_{1}^{1/2}$ & $-3.468385(-1)$   & $-1.425932(9)$ & $-2.066515(11) $ \\ 
$a_{0}^{3/2}$ & $1.016398(-2)$    & $8.148532(10)$ & $ 1.250476(14) $ \\ 
$a_{1}^{3/2}$ & $-6.936930(-1)$   & $-2.929768(9)$ & $ -3.4880108(11) $ \\  
%\hline
%\hline
\end{tabular}
\end{ruledtabular}
\end{table}

%
%
%
%========= FIGURES =================================================== 

\pagebreak 
\
\cleardoublepage
%\newpage 

%
%FIGURE 1
%
%
%===================================================================== 
%
%FIGURE 2
%\pagebreak
% end of file template.tex
%
%===================================================================== 
%
%FIGURE 3
%\pagebreak

%
%===================================================================== 
%
%
%FIGURE 4
%\pagebreak

%
%===================================================================== 

%FIGURE 5
%\pagebreak
%
%===================================================================== 

%FIGURE 6
%\pagebreak

%===================================================================== 

%FIGURE 7
%\pagebreak

%===================================================================== 

%FIGURE 8
%\pagebreak

%===================================================================== 

\end{document}